\documentclass[article,nojss]{jss}

\usepackage{thumbpdf}
\usepackage{amsfonts,amstext,amsmath,amssymb,amsthm}
\usepackage{color}
\usepackage{rotating}
\usepackage{verbatim}
\newcommand{\ie}{\textit{i.e.}~}
\newcommand{\eg}{\textit{e.g.}~}

\renewcommand{\Prob}{\mathbb{P}}
\newcommand{\Ex}{\mathbb{E}}
\newcommand{\RR}{\mathbb{R}}
\newcommand{\eps}{\varepsilon}
\newcommand{\prodname}{tensor }

 \DeclareMathOperator{\Var}{Var}

 \DeclareMathOperator{\diag}{diag}

 \DeclareMathOperator*{\argmin}{{arg\,min}}

\def \bvec {\text{\boldmath$b$}}    
    
    \def \mD {\text{\boldmath$D$}}

    \def \mP {\text{\boldmath$P$}}

\def \xvec {\text{\boldmath$x$}}    \def \mX {\text{\boldmath$X$}}

 \def \calN {\mathcal N}

 \def \calU {\mathcal U}

\def \alphavec        {\text{\boldmath$\alpha$}}
\def \betavec         {\text{\boldmath$\beta$}}
\def \gammavec        {\text{\boldmath$\gamma$}}

\def \YXx {Y | \mX = \xvec}
\def \Yx  {Y_\xvec}
\def \PYX {\Prob_{Y, \mX}}
\def \hatPYX {\hat{\Prob}_{Y, \mX}}
\def \PX {\Prob_{\mX}}

\def \ExYX {\Ex_{Y, \mX}}
\def \hatExYX {\hat{\Ex}_{Y, \mX}}

\def \ExY {\Ex_{Y}}
\def \ExX {\Ex_{\mX}}
\def \PYXx {\Prob_{\YXx}}

\def \sumj {\sum_{j = 1}^J}
\def \sumi {\sum_{i = 1}^N}
\def \sumimath {\sum_{\imath = 1}^n}
\def \one {\mathbf{1}}
\def \jstar {j^\star}

\newtheorem{lem}{Lemma}
\newtheorem{thm}{Theorem}
\newtheorem{corr}{Corollary}

\author{Torsten Hothorn \\ LMU M\"unchen \& \\ Universit\"at Z\"urich \\
   \And Thomas Kneib\\ Universit\"at G\"ottingen \\
   \And Peter B\"uhlmann \\ ETH Z\"urich}
\Plainauthor{Torsten Hothorn, Thomas Kneib, and Peter B\"uhlmann}

\title{Conditional Transformation Models \\ Extended Version$^\star$}
\Plaintitle{Conditional Transformation Models}
\Shorttitle{Conditional Transformation Models}

\Abstract{

The ultimate goal of regression analysis is to obtain information about the
conditional distribution of a response given a set of explanatory variables. 
This goal is, however, seldom achieved because most established regression
models only estimate the conditional mean as a function of the explanatory
variables and assume that higher moments are not affected by the regressors. 
The underlying reason for such a restriction is the assumption of additivity
of signal and noise.  We propose to relax this common assumption in the
framework of transformation models.  The novel class of semiparametric
regression models proposed herein allows transformation functions to depend
on explanatory variables.  These transformation functions are estimated by
regularised optimisation of scoring rules for probabilistic forecasts, \eg 
the continuous ranked probability score.  The corresponding estimated
conditional distribution functions are consistent.  Conditional transformation models
are potentially useful for describing possible heteroscedasticity, comparing spatially
varying distributions, identifying extreme events, deriving prediction
intervals and selecting variables beyond mean regression effects.  An
empirical investigation based on a heteroscedastic varying coefficient
simulation model demonstrates that semiparametric estimation of conditional
distribution functions can be more beneficial than kernel-based
non-parametric approaches or parametric generalised additive models for
location, scale and shape.

}

\Plainkeywords{structured additive regression}

\Address{
  Torsten Hothorn\\
  Institut f\"ur Statistik \hfill ISPM, Abteilung Biostatistik \\
  Ludwig-Maximilians-Universit\"at M\"unchen \hfill Universit\"at Z\"urich \\
  Ludwigstra{\ss}e 33 \hfill Hirschengraben 84\\
  DE-80539 M\"unchen, Germany \hfill CH-8001 Z\"urich, Switzerland \\
  E-mail: \email{Torsten.Hothorn@R-project.org}\\
  URL: \url{http://www.stat.uni-muenchen.de/~hothorn/}\\

  Thomas Kneib \\
  Wirtschaftswissenschaftliche Fakult\"at \\
  Georg-August-Universit\"at G\"ottingen \\
  Platz der G\"ottinger Sieben 5 \\
  DE-37073 G\"ottingen, Germany \\

  Peter B\"uhlmann \\
  Seminar f\"{u}r Statistik \\
  ETH Z\"{u}rich \\
  CH-8092 Z\"{u}rich, Switzerland

}

\begin{document}

\footnote{
$^\star$This is an extended version of
\cite{Hothorn_Kneib_Buehlmann_2013}. Please cite this reference unless
you're referring to the additional applications presented in this document. \\
\texttt{ctm.pdf} compiled \today~from \texttt{ctm.tex} version 5081}

\section{Introduction}

One of the famous ``Top ten reasons to become a statistician'' is that
statisticians are ``mean lovers'' \citep{Friedman_Friedman_Amoo_2002}, referring of course to our obsession with means.
%% Statistics as a profession is obsessed with means.
Whenever a distribution is too complex to think or expound upon, we
focus on the mean as a single real number describing the centre of the
distribution and block out other characteristics such as variance, skewness
and kurtosis.  Our willingness to simplify distributions this way is most
apparent when we deal with many distributions at a time, as in a regression
setting where we describe the conditional distribution $\PYXx$ of a response
$Y \in \RR$ given different configurations of explanatory variables $\mX =
\xvec \in \chi$.  Many regression models focus on the conditional mean
$\Ex(Y | \mX = \xvec)$ and treat higher moments of the conditional
distribution $\PYXx$ as fixed or nuisance parameters that must not depend on the
explanatory variables.  As a consequence, model inference crucially relies
on assumptions such as homoscedasticity and symmetry.  Information on
the scale of the response, for example prediction intervals, derived from
such models also depends on these assumptions.  Here, we propose a
new class of conditional transformation models that allow the conditional
distribution function $\Prob(Y \le \upsilon | \mX = \xvec)$ to be estimated
directly and semiparametrically under rather weak assumptions.  Before we
introduce this class of models in Section~\ref{sec:ctm}, we will attempt to set a
scene of contemporary regression in the light of \cite{Gilchrist_2008} and
place the major players onto this stage.

Let $\Yx = (\YXx) \sim \PYXx$ denote the conditional distribution of
response $Y$ given explanatory variables $\mX = \xvec$.  We assume that
$\PYXx$ is dominated by some measure $\mu$ and has
the conditional distribution function $\Prob(Y \le \upsilon | \mX = \xvec)$.  A
regression model describes the distribution $\PYXx$, or certain
characteristics of it, as a function of the explanatory variables $\xvec$.
We estimate such models based on samples of pairs of random variables $(Y,
\mX)$ from the joint distribution $\PYX$.  It is convenient to assume that a
regression model consists of signal and noise, \ie a deterministic part
and an error term. In the following, we denote the error term by $Q(U)$,
where $U \sim \calU[0, 1]$ is a uniform random variable independent of $\mX$ 
and $Q: \RR \rightarrow \RR$ is the quantile function of an 
absolutely continuous distribution.

Apart from non-parametric kernel estimators of the conditional
distribution function \citep{Hall_Wolff_Yao_1999, Hall_Mueller_2003,
Li_Racine_2008}, there are two common ways to model the influence of the
explanatory variables $\xvec$ on the response $\Yx$:
\begin{eqnarray}
\Yx & = & r(Q(U) | \xvec) \quad \text{``mean or quantile regression models'' and} \label{mod:reg} \\
h(\Yx | \xvec) & = & Q(U) \quad \quad \quad \qquad \text{``transformation models''.} \label{mod:trafo}
\end{eqnarray}
For each $\xvec \in \chi$, the regression function $r(\cdot | \xvec): \RR \rightarrow \RR$
transforms the error term $Q(U)$ in a monotone increasing way. The
inverse regression function $h(\cdot | \xvec) = r^{-1}(\cdot | \xvec): \RR \rightarrow \RR$ is also
monotone increasing. Because $h$ transforms the response, it is known as a
transformation function, and models in the form of (\ref{mod:trafo})
are called transformation models.

A major assumption underlying almost all regression models of class
(\ref{mod:reg}) is that the regression function $r$ is the
sum of the deterministic part $r_\xvec: \chi \rightarrow \RR$, which depends on
the explanatory variables, and the error term:
\begin{eqnarray*}
r(Q(U) | \xvec) = r_\xvec(\xvec) + Q(U).
\end{eqnarray*}
When $\Ex(Q(U)) = 0$, we get $r_\xvec(\xvec) = \Ex(Y | \mX = \xvec)$, \eg
linear or additive models depending on the functional form of $r_\xvec$.
Model inference is commonly based on the normal error assumption, \ie
$Q(U) = \sigma \Phi^{-1}(U)$, where $\sigma > 0$ is a scale parameter and
$\Phi^{-1}(U) \sim \calN(0, 1)$. Linear heteroscedastic regression models
\citep{Carroll_Ruppert_1982} allow describing the variance as a function of
the explanatory variables $Q(U) = \sigma(\xvec) \tilde{Q}(U)$,
where $\sigma(\xvec)$ is a (usually log-linear) function of $\xvec$ and $\tilde{Q}$
is the quantile function of a symmetric distribution
with $\Ex(\tilde{Q}(U)) = 0$. In time series analysis, GARCH models
\citep{Bollerslev_1986} share this idea.
A novel semiparametric approach is extended generalised additive models,
where additive functions of the explanatory variables describe location,
scale and shape (GAMLSS) of a certain parametric conditional distribution
of the response given the explanatory variables
\citep{Rigby_Stasinopoulos_2005}.  If the assumption of a certain parametric
form of the conditional distribution is questionable, $r_\xvec$ describes
the $\tau$ quantile of $\Yx$ when the quantile function $Q$ is such that
$Q(\tau) = 0$ for some $\tau \in (0, 1)$.  This leads us to quantile
regression \citep{Koenker_2005}, which is according to \cite{Stigler_2010} the ``best
approach to robust methods in higher dimensional linear model problems''. 
Estimating the complete conditional quantile function is less
straightforward since we have to fit separate models for a grid of
probabilities $\tau$, and the resulting regression quantiles may cross. 
Solutions to this problem can be obtained by combining all quantile fits in
one joint model based on, for example, location-scale models \citep{He_1997}
or quantile sheets \citep{Schnabel_Eilers_2012}, or by monotonising the estimated
quantile curves using non-decreasing rearrangements
\citep{Dette_Volgushev_2008}.

For transformation models (\ref{mod:trafo}), additivity
is assumed on the scale of the inverse regression function $h$:
\begin{eqnarray*}
h(\Yx | \xvec) = h_Y(\Yx) + h_\xvec(\xvec).
\end{eqnarray*}
When $\Ex(Q(U)) = 0$, we get $-h_\xvec(\xvec) = \Ex(h_Y(\Yx)) = \Ex(h_Y(Y) | \mX =
\xvec)$.  The monotone transformation function $h_Y: \RR \rightarrow \RR$
does not depend on $\xvec$ and might be known in advance (Box-Cox
transformation models with fixed parameters, accelerated failure time
models) or is commonly treated as a nuisance parameter (Cox model,
proportional odds model).  One is usually interested in estimating the
function $h_\xvec: \chi \rightarrow \RR$, which describes the conditional
mean of the \textit{transformed} response $h_Y(\Yx)$.  The class of transformation
models is rich and very actively researched, most prominently in 
literature on the analysis of survival data.  For example, the linear Weibull accelerated failure
time model assumes a log transformation $h_Y(\Yx) = \log(\Yx)$, a linear
function for the conditional mean of the log-transformed response
$h_\xvec(\xvec) = \xvec^\top \alphavec$, and a Weibull-distributed error
term $Q(U) = \sigma Q_\text{Weibull}(U)$.  For the Cox additive model,
$h_Y(\Yx) = \log(\Lambda(\Yx))$ is based on the unspecified integrated
baseline hazard function $\Lambda$, $h_\xvec(\xvec) = \sum_{j = 1}^J
h_{\xvec, j}(\xvec)$ is the sum of $J$ smooth terms depending on the
explanatory variables and $Q(U) = -\log(-\log(U))$ is the quantile function
of the extreme value distribution.  The proportional odds model has
$h_Y(\Yx) = \log(\Gamma(\Yx))$, with $\Gamma$ being an unknown monotone
increasing function, and $Q(U) = \log(U / (1 - U))$ is the quantile function
of the logistic distribution.  \cite{Doksum_Gasko_1990} discussed the
flexibility of this class of models, and \cite{Cheng_Wei_Ying_1995}
introduced a generic algorithm for linear transformation model estimation,
that is, for models with $h_\xvec(\xvec) = \xvec^\top \alphavec$, treating
the transformation function $h_Y$ as a nuisance.

In recent years, transformation models have been extended in two directions.
In the first direction, more flexible forms for the conditional mean function $h_\xvec$ have
been introduced, \eg the partially linear transformation model
$h_\xvec(\xvec) = \xvec^\top (0, \alphavec)^\top + h_\text{smooth}(x_1)$
\citep[where $h_\text{smooth}$ is a smooth function of
the first variable $x_1$;][]{Lu_Zhang_2010}, the varying coefficient model $h_\xvec(\xvec) =
\xvec^\top (0, 0, \alphavec)^\top + h_\text{smooth}(x_1)x_2$
\citep{Chen_Tong_2010}, random effects models \citep{Zeng_Lin_Yin_2005}, and
various approaches to additive transformation and accelerated failure time
models, such as the boosting approaches by \cite{Lu_Li_2008}
and \cite{Schmid_Hothorn_2008}.  In the second direction, a number of authors have considered
algorithms that estimate $h_Y$ and (partially) linear functions $h_\xvec$
simultaneously, usually by a spline expansion of $h_Y$
\citep[\eg][]{Shen_1998, Cheng_Wang_2011}, as an alternative to the common
practise of estimating $h_Y$ \textit{post-hoc} by some non-parametric procedure
such as the Breslow estimator.

Although the transformation function $h_Y$ is typically treated as an infinite
dimensional nuisance parameter, it is important to note that $h_Y$ contains
information about higher moments of $\Yx$, most importantly variance and
skewness.  Simultaneous estimation of $h_Y$ and $h_\xvec$ is therefore
extremely attractive because we can obtain information about the mean and
higher moments of the transformed response at the same time.  However, owing
to the decomposition of the regression function $r$ or the transformation
function $h$ into both a deterministic part depending on the explanatory
variables ($r_\xvec$ or $h_\xvec$) and a random part depending on the
response ($h_Y$) or error term ($Q(U)$), higher moments of the conditional
distribution of $Y$ given $\mX = \xvec$ must not depend on the explanatory
variables in mean regression and transformation models.  As a consequence, the
corresponding models cannot capture heteroscedasticity or skewness that is
induced by certain configurations of the explanatory variables. Therefore,
we cannot detect these potentially interesting patterns, and our models
will perform poorly when probability forecasts, prediction intervals
or other functionals of the conditional distribution are of special interest.

Recently, \cite{Wu_Tian_Yu_2010} proposed a novel transformation model for
longitudinal data that partially addresses this issue. For responses and
explanatory variables $\mX(t)$ observed at time $t$, the model assumes
\begin{eqnarray*}
h(\Yx | t, \xvec) = h_Y(\Yx | t) + \xvec(t)^\top \alphavec(t).
\end{eqnarray*}
Here, the transformation $h_Y$ is conditional on time, and higher moments may
vary with time. However,
since $h_Y$ does not depend on the explanatory variables $\xvec$, these
higher moments may not vary with one or more of the explanatory variables.
In the context of longitudinal data with functional explanatory variables,
\cite{Chen_Mueller_2012} consider a similar model, where the regression
coefficients for functional principle components may depend on time $t$ and
the response $\Yx$. Our contribution is a class of transformation models
where the transformation function is conditional on the explanatory
variables in the sense that the transformation function, and therefore
higher moments of the conditional distribution of the response,
may depend on potentially all explanatory variables. As a consequence,
the models suggested here are able to deal with heteroscedasticity and
skewness that can be regressed on the explanatory variables, and we will
show that reliable estimates of the complete conditional
distribution function and functionals thereof can be obtained.

We will introduce the conditional transformation models
(Section~\ref{sec:ctm}), discuss the underlying model assumptions, and embed
the estimation problem in the empirical risk minimisation framework
(Section~\ref{sec:est}).  For the sake of simplicity, we restrict ourselves
to continuous responses $Y$ that have been observed without
censoring.  Similar to other transformation models, conditional
transformation models are flexible enough to deal with discrete responses
and survival times, as will be discussed in later sections.
%% Sections~\ref{sec:app} and \ref{sec:dis}.
%% but we want to avoid the additional complex notation.  
We present a computationally efficient algorithm for fitting the models in
Section~\ref{sec:boost}.  We study the asymptotic properties of the
estimated conditional distribution functions in Section~\ref{sec:con}. 
The practical benefits of modelling the influence of explanatory variables
on the variance and higher moments of the response' distribution are
demonstrated in Section~\ref{sec:app} with a special emphasis on
distributional characteristics of childhood nutrition in India and on 
prediction intervals for birth weights of small foetus.
Finally, we use a heteroscedastic varying
coefficient simulation model to evaluate the empirical performance of the
proposed algorithm and compare the quality of conditional distribution
functions estimated by a conditional transformation model and established
parametric and non-parametric procedures in Section~\ref{sec:eval}.

\section{Conditional Transformation Models} \label{sec:ctm}

An attractive feature of transformation models is their close connection to
the conditional distribution function. With the transformation function
$h(\Yx | \xvec) = Q(U)$, one can evaluate the conditional distribution function
of response $Y$ given the explanatory variables $\xvec$ via
\begin{eqnarray*}
\Prob(Y \le \upsilon | \mX = \xvec) = \Prob(h(Y|\xvec) \le h(\upsilon|\xvec)) =
F(h(\upsilon | \xvec))
\end{eqnarray*}
with absolute continuous distribution function $F = Q^{-1}$. For additive
transformation functions $h = h_Y + h_\xvec$, the conditional distribution
function reads $F(h(\upsilon | \xvec)) = F(h_Y(\upsilon) + h_\xvec(\xvec))$,
\ie the distribution is evaluated for a transformed and shifted version of
$Y$.  Higher moments only depend on the transformation $h_Y$ and thus cannot
be influenced by the explanatory variables.  Consequently, one has to
avoid the additivity in the model $h = h_Y + h_\xvec$ to allow the explanatory
variables to impact also higher moments.  We therefore suggest a novel
transformation model based on an alternative additive decomposition of the
transformation function $h$ into $J$ partial transformation functions
for all $\xvec \in \chi$:
\begin{eqnarray} \label{mod:ctm}
h(\upsilon | \xvec) = \sumj h_j(\upsilon | \xvec),
\end{eqnarray}
where $h(\upsilon | \xvec)$ is the monotone transformation function of $\upsilon$.
In this model, the transformation function $h(\Yx | \xvec)$ and 
the partial transformation functions $h_j(\cdot | \xvec): \RR \rightarrow \RR$
are conditional on $\xvec$ in the sense that not only the mean of 
$\Yx$ depends on the explanatory variables.
For this reason, we coin models of the form (\ref{mod:ctm})
\emph{Conditional Transformation Models} (CTMs).
Clearly, model (\ref{mod:ctm}) imposes an assumption, namely additivity of the
conditional distribution function on the scale of the
quantile function $Q$:
\begin{eqnarray*}
Q(\Prob(Y \le \upsilon | \mX = \xvec)) = \sumj h_j(\upsilon | \xvec).
\end{eqnarray*}
It should be noted that here we assume additivity of the transformation
function $h$ and not additivity on the scale of the regression function $r$
as it is common for additive mean or quantile regression models
(\ref{mod:reg}). Furthermore, monotonicity of $h_j$ is sufficient but not
necessary for $h$ being monotone. 
Of course, we have to make further assumptions on $h_j$ to obtain
reasonable models, but these assumptions are problem specific, and we will
therefore postpone these issues until Section~\ref{sec:app}. 
To ensure identifiability, we assume without loss of generality
that the partial transformation functions are centred around zero
$\ExY \ExX h_j(Y | \mX) = 0$ for all $j = 1, \dots, J$ for 
non-systematic error terms ($\Ex(Q(U)) = 0$).

%%% cause-specific hazard function?

\section{Estimating Conditional Transformation Models} \label{sec:est}

The estimation of conditional distribution functions can be reformulated as
a mean regression problem since $\Prob(Y \le \upsilon | \mX = \xvec) =
\Ex(I(Y \le \upsilon) | \mX = \xvec)$ for the binary event $Y \le \upsilon$;
this connection is widely used \citep[\eg
by][]{Hall_Mueller_2003,Chen_Mueller_2012}.  Similar to the approach of
fitting multiple quantile regression models to obtain an estimate of the
conditional quantile function, one could estimate the models $\Ex(I(Y \le
\upsilon) | \mX = \xvec)$ for a grid of $\upsilon$ values separately.
However, we instead suggest estimating conditional transformation models by the
application of an integrated loss function that allows the whole
conditional distribution function to be obtained in one step.

Let $\rho$ denote a function of measuring the loss of the probability
$F(h(\upsilon | \mX))$ for the binary event $Y \le \upsilon$.
One candidate loss function is
\begin{eqnarray*}
\rho_\text{bin}((Y \le \upsilon, \mX), h(\upsilon | \mX))
  & := & -[I(Y \le \upsilon) \log\{F(h(\upsilon | \mX))\} + \\
  & & \quad  \{1 - I(Y \le \upsilon)\} \log\{1 - F(h(\upsilon | \mX))\}] \ge 0,
\end{eqnarray*}
the negative log-likelihood of the binomial model
$(Y \le \upsilon | \mX = \xvec) \sim B(1, F(h(\upsilon | \xvec)))$
for the binary event $Y \le \upsilon$ with link function $Q = F^{-1}$.
Alternatively, one may consider the squared or absolute error losses
\begin{eqnarray*}
\rho_\text{sqe}((Y \le \upsilon, \mX), h(\upsilon | \mX))
  & := & \frac{1}{2}|I(Y \le \upsilon) - F(h(\upsilon | \mX))|^2 \ge 0 \\
\rho_\text{abe}((Y \le \upsilon, \mX), h(\upsilon | \mX))
  & := & |I(Y \le \upsilon) - F(h(\upsilon | \mX))| \ge 0.
\end{eqnarray*}
The squared error loss $\rho_\text{sqe}$ is also known as the Brier score, and
the absolute loss $\rho_\text{abe}$ has been applied for assessing survival probabilities in the
Cox model by \cite{Schemper_Henderson_2000}.  We define the loss function $\ell$ for
estimating conditional transformation models as integrated loss $\rho$
with respect to a measure $\mu$ dominating the conditional distribution
$\Prob(Y \le \upsilon | \mX = \xvec)$:
\begin{eqnarray*}
\ell((Y, \mX), h) :=
  \int \rho((Y \le \upsilon, \mX), h(\upsilon | \mX)) \,d\mu(\upsilon) \ge 0.
\end{eqnarray*}
In the context of scoring rules, the loss $\ell$ based on
$\rho_\text{sqe}$ is known as the continuous
ranked probability score (CPRS) or integrated Brier score and is a
proper scoring rule for assessing the quality of probabilistic
or distributional forecasts \citep[see][for an overview]{Gneiting_Raftery_2007}.
It seems natural to apply these scores as loss functions for model
estimation, but we are only aware of the work of \cite{Gneiting_Raftery_Westveld_2005},
who directly optimise the CPRS for estimating Gaussian predictive probability
density functions for continuous weather variables. In the context of non-parametric
or semiparametric estimation of conditional distribution functions,
minimisation of the empirical analog of the risk function
\begin{eqnarray*}
\ExYX \ell((Y, \mX), h) =
  \int \int \rho((y \le \upsilon, \xvec), h(\upsilon | \xvec))
    \,d\mu(\upsilon) \,d \PYX(y, \xvec) \ge 0
\end{eqnarray*}
for estimating conditional distribution functions has not yet been considered.

Model estimation based on the risk $\ExYX \ell((Y, \mX), h)$ is reasonable because
the corresponding optimisation problem is convex and attains its minimum for
the true conditional transformation function $h$. We summarise these facts
in the following lemma, whose proof in given in the Appendix.
\begin{lem}
The risk $\ExYX \ell((Y, \mX), h)$ is convex in $h$ for convex losses
$\rho$ in $h$.  The population minimiser of $\ExYX \ell((Y, \mX), h)$ for $\rho =
\rho_\text{bin}$ and $\rho = \rho_\text{sqe}$ is $h(\upsilon | \xvec) = Q(\Prob(Y \le \upsilon
| \mX = \xvec))$.  For $\rho = \rho_\text{abe}$, the minimiser is 
\begin{eqnarray*}
h(\upsilon | \xvec) = \left\{ \begin{array}{rl} 
-\infty: & \Prob(Y \le \upsilon | \mX = \xvec) \le 0.5 \\
\infty: & \Prob(Y \le \upsilon | \mX = \xvec) > 0.5.
\end{array} \right.
\end{eqnarray*}
\end{lem}

The corresponding empirical risk function defined by the data is
\begin{eqnarray*}
\hatExYX \ell((Y, \mX), f) =
  \int \int \rho((y \le \upsilon, \xvec), h(\upsilon | \xvec))
  \,d\mu(\upsilon) \,d \hatPYX(y, \xvec) \ge 0.
\end{eqnarray*}
Based on an i.i.d.~random sample $(Y_i, \mX_i) \sim \PYX, i = 1, \dots, N$
of $N$ observations from the joint distribution of response and explanatory
variables, we define $\hatPYX$ as the distribution putting mass $w_i > 0$ on
observation $i$ ($w_i \equiv N^{-1}$ for the empirical distribution).
For computational convenience, we also approximate the measure $\mu$ by
the discrete uniform measure $\hat{\mu}$, 
which puts mass $n^{-1}$ on each element of the equi-distant grid
$\upsilon_1 < \dots < \upsilon_n \in \RR$ over the response space. The
number of grid points $n$ has to be sufficiently large to closely approximate the
integral. The empirical risk is then
\begin{eqnarray} \label{mod:risk}
\hatExYX \ell((Y, \mX), h) & = &
  \sumi w_i n^{-1} \sumimath
  \rho((Y_i \le \upsilon_\imath, \mX_i), h(\upsilon_\imath | \mX_i)) \\ \nonumber
  & = & n^{-1}
  \sumi \sumimath w_i \rho((Y_i \le \upsilon_\imath, \mX_i),
                            h(\upsilon_\imath | \mX_i)).
\end{eqnarray}
This risk is the weighted empirical risk for loss function $\rho$
evaluated at the observations $(Y_i \le \upsilon_\imath, \mX_i)$ for $i = 1,
\dots, N$ and $\imath = 1, \dots, n$.  Consequently, we can apply algorithms
for fitting generalised additive models to the binary responses $Y_i \le
\upsilon_\imath$ under loss $\rho$ for estimating model (\ref{mod:ctm}).
Although this seems to be rather straightforward, there are two issues to
consider.  First, simply expanding the observations over the grid
$\upsilon_1 < \dots < \upsilon_n$ increases the computational complexity by
$n$, which, even for moderately large sample sizes $N$, renders computing and
storage rather burdensome.  Second, unconstrained minimisation of the
empirical risk, \ie no smoothness of $h$ in its first argument and $h$
being independent of the conditioning $\xvec$, leads to estimates
$F(\hat{h}(\upsilon | \xvec)) = \hat{\Prob}(Y \le \upsilon) = N^{-1} \sum_{i
= 1}^N I(Y_i \le \upsilon)$,  \ie the empirical cumulative distribution
function of $Y$ for $\rho_\text{bin}$ and $\rho_\text{sqe}$ with $w_i =
N^{-1}$.  For $\rho_\text{abe}$, the empirical risk is minimised by
$F(\hat{h}(\upsilon | \xvec)) = 0$ for all $\upsilon$ with $\hat{\Prob}(Y
\le \upsilon) < 0.5$ and otherwise by $F(\hat{h}(\upsilon | \xvec)) = 1$.
Therefore, careful regularisation is absolutely necessary to obtain
reasonable models that lead to smooth conditional distribution functions
(\ie smoothing in the $Y$-direction) and that are similar for similar
configurations of the explanatory variables (\ie smoothing in the $\mX$-direction).
Instead of adding a direct penalisation term to the empirical risk, we
propose in the next section a boosting algorithm for empirical risk minimisation that indirectly
controls the functional form and complexity of the estimate $\hat{h}$.

%% Achtung: Risk ist dann d P_Y und nicht d mu !
%%Alternatively, we could have used the marginal empirical distribution
%%$\hat{\mu} = \hat{Prob}_Y$ that puts mass $N^{-1}$ on
%%the observations $Y_1, \dots, Y_N$ and obtain the V-statistic
%%\begin{eqnarray*}
%%\hatExYX \ell((Y, \mX), h) =
%%  N^{-2} \sumi \sum_{\imath = 1}^N
%%  \rho((Y_i \le Y_\imath, \mX_i), h(Y_\imath | \mX_i)).
%%\end{eqnarray*}

\section{Boosting Conditional Transformation Models} \label{sec:boost}

We propose to fit conditional transformation models~(\ref{mod:ctm}) by a
variant of component-wise boosting for minimising the empirical risk
(\ref{mod:risk}) with penalisation.  In this class of algorithms,
regularisation is achieved indirectly via the application of penalised
base-learners and the complexity of the whole model is controlled by the
number of boosting iterations.  We refer the reader to
\cite{Buehlmann_Hothorn_2007} for a detailed introduction to component-wise
boosting.

For conditional transformation models, we parameterise the partial transformation
functions for all $j = 1, \dots, J$ as
\begin{eqnarray} \label{mod:partial}
h_j(\upsilon | \xvec) = \left(\bvec_j(\xvec)^\top \otimes
                              \bvec_0(\upsilon)^\top\right) \gammavec_j \in
\RR, \qquad \gammavec_j \in \RR^{K_jK_0},
\end{eqnarray}
where $\bvec_j(\xvec)^\top \otimes \bvec_0(\upsilon)^\top$ denotes the
\prodname product of two sets of basis functions $\bvec_j : \chi \rightarrow
\RR^{K_j}$ and $\bvec_0 : \RR \rightarrow \RR^{K_0}$.  Here, $\bvec_0$ is a
basis along the grid of $\upsilon$ values that determines the functional
form of the response transformation.  The basis $\bvec_j$ defines how this
transformation may vary with certain aspects of the explanatory variables.
The \prodname product may be interpreted as a generalised interaction effect
(further illustrated in Section~\ref{sec:app}).  For each partial transformation
function $h_j$, we typically want to obtain an estimate that is smooth in its
first argument $\upsilon$ and smooth in the conditioning variable $\xvec$.
Therefore, the bases are supplemented with appropriate, pre-specified penalty
matrices $\mP_j \in \RR^{K_j \times K_j}$ and $\mP_0 \in \RR^{K_0 \times
K_0}$, inducing the penalty matrix $\mP_{0j} = (\lambda_0 \mP_j \otimes
\one_{K_0} + \lambda_j \one_{K_j} \otimes \mP_0)$ with smoothing parameters
$\lambda_0\ge0$ and $\lambda_j\ge0$ for the \prodname product basis.  The
base-learners corresponding to the partial transformation functions fitted
to the negative gradients in each iteration of the boosting algorithm are
then Ridge-type linear models with penalty matrix $\mP_{0j}$.  In more
detail, we apply the following algorithm for fitting conditional
transformation models with partial transformation functions of the form
(\ref{mod:partial}):

%%<FIXME> \one_{K_0} meint die Einheitsmatrix, besser \mathbf{I}_{K_0}
%% + Erklaerung </FIXME>

\paragraph{Algorithm: Boosting for Conditional Transformation Models}

\begin{description}
\item{(Init)}
Initialise the parameters $\gammavec^{[0]}_j \equiv 0$ for $j = 1, \dots,
J$, the step-size $\nu \in (0, 1)$ and the smoothing parameters
$\lambda_j, j = 0, \dots, J$. Define the grid
$\upsilon_1 < Y_{(1)} < \dots < Y_{(N)} \le \upsilon_n$. Set $m := 0$.

\item{(Gradient)}

Compute the negative gradient:
\begin{eqnarray*}
U_{i\imath} := - \left. \frac{\partial}{\partial h}
  \rho((Y_i \le \upsilon_\imath, \mX_i), h)
  \right|_{h = \hat{h}^{[m]}_{i\imath}}
\end{eqnarray*}
with $\hat{h}^{[m]}_{i\imath} = \sumj
\left(\bvec_j(\mX_i)^\top \otimes \bvec_0(\upsilon_\imath)^\top\right)
\gammavec_j^{[m]}$.

Fit the base-learners for $j = 1, \dots, J$:
\begin{eqnarray} \label{mod:base}
\hat{\betavec}_j = \argmin_{\betavec \in \RR^{K_jK_0}}
  \sumi \sumimath w_i \left\{U_{i\imath} - \left(\bvec_j(\mX_i)^\top \otimes
    \bvec_0(\upsilon_\imath)^\top\right)\betavec\right\}^2
    + \betavec^\top \mP_{0j} \betavec
\end{eqnarray}
with penalty matrix $\mP_{0j}$.

Select the best base-learner:
\begin{eqnarray*}
j^\star = \argmin_{j = 1, \dots, J} \sumi \sumimath w_i
  \left\{U_{i\imath} - \left(\bvec_j(\mX_i)^\top \otimes
         \bvec_0(\upsilon_\imath)^\top\right)\hat{\betavec}_j\right\}^2
\end{eqnarray*}
\item{(Update)} the parameters
$\gammavec_{\jstar}^{[m + 1]} = \gammavec_{\jstar}^{[m]} +
                                \nu \hat{\betavec}_{\jstar}$ and keep all other parameters fixed,
\ie $\gammavec_{j}^{[m + 1]} = \gammavec_{j}^{[m]}$, $j\neq \jstar$.
\item{Iterate (Gradient) and (Update)} 
\item{(Stop)} if $m = M$. Output the final model
\begin{eqnarray*}
\hat{\Prob}(Y \le \upsilon | \mX = \xvec) =
  F\left(\hat{h}^{[M]}\left(\upsilon | \xvec\right)\right)
= F\left(\sumj \left(\bvec_j(\xvec)^\top \otimes
                       \bvec_0(\upsilon)^\top\right)\gammavec^{[M]}_j\right)
\end{eqnarray*}
as a function of arbitrary $\upsilon \in \RR$ and $\xvec \in \chi$.
\end{description}

Before we investigate the asymptotic properties of the resulting estimates,
we will discuss some details of this generic algorithm in the following.

\paragraph{Model Specification.}
The basis functions $\bvec_0$ and $\bvec_j$ determine the form of the fitted
model, and their choice is problem specific. In the simplest situation, in which
the conditional distribution of $Y$ given only one numeric explanatory
variable $x_1$ shall be estimated, one could use the basis functions
$\bvec_0(\upsilon) = (1, \upsilon)^\top$ and $\bvec_1(\xvec) = (1, x_1)^\top$.
The corresponding base-learner is then defined by the linear function
\begin{eqnarray*}
\left((1, x_1) \otimes (1, \upsilon)\right) \gammavec_1 =
  (1, \upsilon, x_1, x_1 \upsilon) \gammavec_1.
\end{eqnarray*}
For each $x_1$, the transformation is linear in $\upsilon$ with intercept
$\gamma_1 + \gamma_3 x_1$ and slope $\gamma_2 + \gamma_4 x_1$, \ie not only
the mean may depend on $x_1$ but also the variance.  Restricting, for example,
$\bvec_0(\upsilon)$ to be constant, \ie $\bvec_0(\upsilon)\equiv1$, allows
the effects of explanatory variables to be restricted to the mean alone.  Assuming
$\bvec_1(\xvec)\equiv1$, on the other hand, yields a transformation function
that is not affected by any explanatory variable.  More flexible basis functions, \eg
$B$-spline basis functions, allow also for higher moments to depend
on the explanatory variables.  We illustrate appropriate choices of basis functions 
in Section~\ref{sec:app}.

\paragraph{Computational Complexity.}
For the estimation of base-learner parameters $\betavec_j$ in
(\ref{mod:base}), it is not necessary to evaluate the Kronecker product
$\otimes$ in (\ref{mod:partial}) and
to compute the $nN \times K_0K_j$ design matrix for the $j$th base-learner.
The base-learners used here are a special form of multidimensional smooth
linear array models \citep{Currie_Durban_Eilers_2006}, where efficient
algorithms for computing Ridge estimates (\ref{mod:base}) exist.
The number of multiplications required for fitting the $j$th base-learner is
approximately $c^6 / (c^2 / N - 1)$, instead of $N^2 c^4$ for the simplest
case with $c = K_0 = K_j$ and $N = n$ 
\citep[see Table~2 in][]{Currie_Durban_Eilers_2006}, and the required memory for storing the
design matrices is of the order $NK_j + NK_0$, instead of $N n K_j K_0$.  Note
that only the gradient vector is of length $N n$; all other objects can be
stored in vectors or matrices growing with either $N$ or $n$, and
an explicit expansion of the observations $(Y_i \le \upsilon_\imath, \mX_i)$
for $i = 1, \dots, N$ and $\imath = 1, \dots, n$ is not necessary.

\paragraph{Choice of Tuning Parameters.}
The number of boosting iterations $M$ is the most important tuning parameter
determined by resampling, \eg by $k$-fold cross-validation
or bootstrapping. For the latter resampling scheme, the weights $w_i$ 
in (\ref{mod:risk}) are
drawn from an $N$-dimensional multinomial distribution with constant
probability parameters $p_i \equiv N^{-1}, i = 1, \dots, N$. The out-of-bootstrap (OOB) empirical risk
with weights $w_i^\text{OOB} = I(w_i = 0)$ is then used as a measure to assess
the quality of the distributional forecasts for a varying number of
boosting iterations $M$. It should be noted that the loss function used to
fit the models is the same function that is used as a scoring rule to assess 
the quality of the probabilistic forecasts of the OOB observations.

The smoothing parameters $\lambda_j, j = 0, \dots, J$ in the penalty
matrices are not tuned but
rather defined such that the $j$th base-learner has low degrees of freedom. For
our computations, we simplified the penalty term to
$\mP_{0j} =  \lambda_j (\mP_j \otimes \one_{K_0} + \one_{K_j} \otimes
\mP_0)$, \ie one parameter controls the smoothness in both directions.
Following \cite{Hofner_Hothorn_Kneib_2011}, the parameters $\lambda_j$
were defined such that each base-learner has the same overall
low degree of freedom. Note that the degree of freedom of the estimated
partial transformation function adapts to the complexity inherent in the
data via the number of boosting iterations $M$ 
\citep{Buehlmann_Yu_2003}. Different smoothness in the two directions can be
imposed by choosing different basis functions for $\bvec_0$ and $\bvec_j$,
\eg a linear basis function for $\bvec_j$ and $B$-splines for
$\bvec_0$.

Other parameters, such as knots or degrees of basis functions or the number
$n$ of grid points the integrated loss function $\ell$ is approximated with
are not considered as tuning parameters. The resulting estimates are rather
insensitive to their different choices. Also, we do not consider the distribution
function $F$ or the loss function $\rho$
as tuning parameter but assume that these are part of the model
specification. Different versions of $F$ and $\rho$ lead to different
negative gradients; these are given in the Appendix.

\paragraph{Monotonicity.}
The resulting estimate $\hat{h}^{[M]}\left(\upsilon | \xvec\right)$ is not
automatically monotone in its first argument. Monotonicity and smoothness in
the $Y$-direction depend on each other, and too-complex estimates tend to suffer
from non-monotonicity. Empirically, based on experiments reported in
Sections~\ref{sec:app} and \ref{sec:eval}, non-monotonicity is a problem in
poorly-fitting models, due to either misspecification, overfitting, or a low
signal-to-noise ratio. From our point of view, inspecting the model for
non-monotonicity is helpful for model diagnostics and can be dealt with by
reducing model complexity. Alternatively, there are three possible
modifications to the algorithm that can be implemented to enforce
monotonicity: (i) fit base-learners under monotonicity constraints
in (\ref{mod:base}), \eg by using the iterative re-penalisation
suggested by \cite{Eilers_2005} and applied to boosting by
\cite{Hofner_Mueller_Hothorn_2011}, (ii) check monotonicity
for each base-learner and select the best among the monotone candidates only, or
(iii) select the base-learner such that it is the best one among all candidates
that lead to monotone updates in $h^{[m]}$. None of these approaches had to be
used for our empirical studies, in which all resulting estimates were monotone
for the appropriate number of boosting iterations $M$.

%\TODO{auch moeglich: $h_j(v | \xvec) = \int_{-\infty}^v \exp(g_j(y | \xvec))
%dy$}
%% reparameterization see chang wang

\paragraph{Model Diagnostics and Overfitting.}
Another convenient feature of transformation models is that, with the
correct model $h$ for absolute continuous random variables $Y$,
the errors $E_i = h(Y_i | \mX_i), i = 1, \dots, N$ are distributed
according to $F$. Therefore, if the observed residuals
$\hat{E}^{[M]}_i = \hat{h}^{[M]}(Y_i | \mX_i)$ are unlikely to come from
distribution $F$, \eg assessed using quantile-quantile plots or
a Kolmogorov-Smirnov statistic, the model is likely to fit the data 
poorly. However, a good agreement between $\hat{E}^{[M]}_i$ and $F$ does not
necessarily mean that the explanatory variables describe the response well.
A high correlation between the ranking of the residuals and the ranking
of the responses $Y_1, \dots, Y_N$ means that the estimated conditional distribution
is very close to the unconditional empirical distribution of the responses.
In this case, either the model may overfit or the response may be
independent of the explanatory variables.

The fitted model may also be used to draw novel responses for given
explanatory variables using the model-based bootstrap via
\begin{eqnarray*}
\tilde{Y}_i = \left\{\upsilon : Q(U_i) = \hat{h}^{[M]}(\upsilon|\mX_i)\right\},
              \quad i = 1, \dots, N
\end{eqnarray*}
where $U_1, \dots, U_N$ are i.i.d.~uniform random variables. The stability
of the model can now be investigated by refitting the model with observations
$(\tilde{Y}_i, \mX_i), i = 1, \dots, N$.

\section{Consistency of Boosted Conditional Transformation Models}
\label{sec:con}

The boosting algorithm presented here is a variant of $L_2$WCBoost
\citep{Buehlmann_Yu_2003} applied to dependent observations with more general
base-learners.  In this section, we will develop a consistency result for the
squared error loss $\rho_\text{sqe}$.  For simplicity, we consider the case
in which the procedure is used with $F(h) = h$ as the identity function,
\ie the error term is uniformly distributed.
Thus, we consider conditional transformation models of the
form
\begin{eqnarray*}
\Prob(Y \le \upsilon | \mX = \xvec) = h(\upsilon|\xvec) = \sum_{j=1}^J
h_j(\upsilon|x_j),
\end{eqnarray*}
where the partial transformation function $h_j$ is conditional on the $j$th
explanatory variable in $\xvec = (x_1, \dots, x_J) \in \chi$ and $\ExY(
N^{-1} \sumi h(Y | \mX_i)) = 1/2$.  Our analysis is for the fixed design case with
deterministic explanatory variables $\mX_i$ or when conditioning on all
$\mX_i$s.  A modification for the random design case could be pursued along
arguments similar to those for $L_2$Boosting as in \citet{Buehlmann_2006}.

As in Section~\ref{sec:boost}, we use a basis expansion of $h(\upsilon|\xvec)$:
\begin{eqnarray*}
h_{N,\gammavec}(\upsilon|\xvec) = \sum_{j=1}^J
 \left(\bvec_j(x_j)^\top \otimes
                              \bvec_0(\upsilon)^\top\right) \gammavec_j =
\sum_{j=1}^J \sum_{k_0 = 1}^{K_{0,N}} \sum_{k_1 = 1}^{K_{1,N}}
\gamma_{j, k_0, k_1} b_{0,k_0}(v) b_{j,k_1}(x_j),
\end{eqnarray*}
where for the sake of simplicity
the number of basis functions $K_{1, N}$ is equal for all $x_j$.
%where $k = (k_1,k_2)$ is a multi-index.

Consider the (empirical) risk functions
\begin{eqnarray*}
R_{n,N} (h) = (nN)^{-1} \sumi \sumimath (I(Y_i \le \upsilon_\imath) -
h(\upsilon_\imath|\mX_i))^2
\end{eqnarray*}
and
\begin{eqnarray*}
R_{n,N,\Ex}(h) = (nN)^{-1} \sumi \sumimath \Ex[(I(Y_i \le \upsilon_\imath)
                           - h(\upsilon_\imath|\mX_i))^2].
\end{eqnarray*}
Denote the projected parameter by
\begin{eqnarray}\label{proj}
\gammavec_{0,N} = \argmin_{\gammavec} R_{n,N,\Ex}(h_{N,\gammavec}).
\end{eqnarray}

%Then
%\begin{eqnarray*}
%\eps_{ir} = I(Y_i \le v_r) - h_{N,\gamma_{0,N}}(v_r|X_i)
%\end{eqnarray*}
%satisfies
%\begin{eqnarray*}
%n^{-1} \sum_{r=1} \Ex[\eps_{ir} b_{0,k_1}(v_r) b_{j,k_2}(X_i^{(j)})]
%= 0\ \mbox{for all},\ k_1,\ k_2.
%\end{eqnarray*}

We make the following assumptions:
\begin{description}
\item[(A1)] The coefficient vector $\gammavec_{0,N}$ is sparse and satisfies
\begin{eqnarray*}
\|\gammavec_{0,N}\|_1 = o\left(\sqrt{\frac{N}{\log(J_N K_{0,N}
K_{1,N})}}\right)\ (N \to \infty).
\end{eqnarray*}
Thereby, the dimensionality  $J = J_N$ can grow with
$N$.
\item[(A2)] The basis functions satisfy: for some $0 < C < \infty$,
\begin{eqnarray*}
\|b_{0,k_0}\|_{\infty} \le C,\ \|b_{j,k_1}\|_{\infty} \le C\ \forall
j, k_0, k_1.
\end{eqnarray*}
\item[(A3)] $$(nN)^{-1} \sumi \sumimath
h_{\gammavec_{0,N}}(\upsilon_\imath|\mX_i)^2 \le D <
  \infty\ \forall n,N$$.
%
%\left(\sum_{j=1}^J \sum_{k_1=1}^{K_{0,N}}
%  \sum_{k_2=1}^{K_{1,N}} \gammavec_{0,j,k} b_{0,k_1}(v_r)
%  b_{j,k_2}(X) \right)^2] \le D < \infty$.
%
%
%$n^{-1} \sum_{r=1}^n \Ex[\left(\sum_{j=1}^J \sum_{k_1=1}^{K_{0,N}}
%  \sum_{k_2=1}^{K_{1,N}} \gammavec_{0,j,k} b_{0,k_1}(v_r)
%  b_{j,k_2}(X) \right)^2] \le D < \infty$.
\end{description}
Assumption (A1) is an $\ell_1$-norm sparsity assumption, (A2) is a mild
restriction since we are modeling $I(Y \le \upsilon)$, and (A3) 
requires that the signal strength does not diverge as $n,N \to
\infty$.

\begin{thm}\label{th1}
Assume (A1)-(A3). Then, for fixed $n$ or for $n = n_N \to \infty\ (N
\to \infty)$, and for $M = M_N \to \infty\ (N \to \infty)$, $M_N =
o(\sqrt{N/\log(J_N K_{0,N} K_{1,N})})$:
\begin{eqnarray*}
(nN)^{-1} \sumi \sumimath (h_{\hat{\gammavec}^{[M]}}(\upsilon_\imath|\mX_i) -
h_{\gammavec_{0,N}}(\upsilon_\imath|\mX_i) )^2 = o_P(1)\ (N \to \infty).
\end{eqnarray*}
\end{thm}
A proof is given in the Appendix.

Convergence of $h_{\gammavec_{0,N}}(\upsilon|\xvec)$ to the true function
$h(\upsilon|\xvec)$ involves approximation theory to achieve
\begin{eqnarray}\label{add1}
(nN)^{-1} \sumi \sumimath (	h_{\gammavec_{0,N}}(\upsilon_\imath|\mX_i) -
h(\upsilon_\imath|\mX_i) )^2 = o(1)\ (n,N \to \infty).
\end{eqnarray}
We typically would want to estimate the function $h(\upsilon|\xvec)$ well
over the whole domain, \eg $[a_\upsilon, b_\upsilon] \times \chi$.  This
may be too ambitious if $J = \text{dim}(\chi) = J_N$ grows with $N$.
Hence, we restrict ourselves to the setting where the number of active variables
$J_{\mathrm{act}} < \infty$ is fixed (from the active set $S$), \ie
\begin{eqnarray*}
h(\upsilon| \xvec) = \sum_{j \in S} h_j(\upsilon| x_j),\ S \subseteq \{1,\ldots ,J\}\
\mbox{with}\ |S| = J_{\mathrm{act}}.
\end{eqnarray*}

For the approximation, we typically would need $K_{0,N},\ K_{1,N} \to
\infty\ (N \to \infty)$ for suitable basis functions and $n = n_N \to
\infty\ (N \to \infty)$; furthermore, the grid $\upsilon_1 < \upsilon_2 <
\ldots < \upsilon_n$ should become dense as $n = n_N \to \infty$, and also
the values $\mX_1^{\mathrm{act}},\ldots ,\mX_N^{\mathrm{act}}$ should become
dense in $\chi_S \subseteq \chi$ as $N \to \infty$ (here,
$\mX^{\mathrm{act}} = \{X_j;\ j \in S\} \in \chi_S$).  If $J = J_N$ grows, but the
number of active variables in the model $J_{\mathrm{act}} < \infty$ is
fixed, then some uniform approximation $h_{\gammavec_{0,N}}(\upsilon|\xvec)
\to h(\upsilon|\xvec)$ is possible under regularity conditions.
%
%For example, (\ref{add1}) holds if $J_{\mathrm{act}} < \infty$ and
%the functions are defined on a compact
%domain, say $[a_v,b_v] \times {\cal X}^J$, if they are continuous and
%if the basis functions are dense.

We provide a summary for a typical situation.
\begin{corr}
Consider the setting as in Theorem \ref{th1}, with $J = J_N$ potentially
growing but fixed dimensionality of the active variables $J_{\mathrm{act}} <
\infty$, $n = n_N \to \infty\ (N \to \infty)$, and the functions are
sufficiently regular such that (\ref{add1}) holds.  Then, for $M = M_N$ as
in Theorem \ref{th1}:
\begin{eqnarray*}
(nN)^{-1} \sumi \sumimath (h_{\hat{\gammavec}^{[M]}}(\upsilon_\imath|\mX_i) -
h(\upsilon_\imath|\mX_i))^2 =  o_P(1)\ (n,N \to \infty).
\end{eqnarray*}
\end{corr}
This result states that the estimated $h_{\hat{\gammavec}^{[M]}}$ are consistent 
for the true transformation function~$h$.

\section{Applications} \label{sec:app}

In this section, we present analyses with special emphasis on higher
moments of the conditional distribution, which have received less attention in
previous analyses of these problems. We show that semiparametric regression
using conditional transformation models is a valuable tool for detecting 
interesting patterns beyond the conditional mean.

\subsection*{``Evolution Canyon'' Bacteria}

The \textit{Bacillus~simplex} populations from ``Evolution Canyons'' I and II in
Israel have recently developed into a model study of bacterial adaptation and
speciation under heterogeneous environmental conditions
\citep{Sikorski_Nevo_2005}.  These two canyons represent similar ecological sites,
$40$ km apart, in which the orientation of the sun yields both a
strongly sun-exposed, hot, `African', south-facing slope and a 
cooler, mesic, lush, `European', north-facing slope within a distance of
only $50$--$400$ m.  Based on DNA sequences, the
\textit{B.~simplex} population phylogenetically splits into two major groups, $\text{GL}_1$
and $\text{GL}_2$.  These main groups are further
subdivided into phylogenetic groups, the so-called `putative ecotypes' (PE),
which show a clear preference for one of the two slope types \citep{Sikorski_Nevo_2005,
Sikorski_Pukall_Stackebrandt_2008}.  $\text{GL}_2$ is composed of only $\text{PE}_1$ and
$\text{PE}_2$; $\text{GL}_1$ consists of 
$\text{PE}_3$--$\text{PE}_9$.  

\cite{Sikorski_Pukall_Stackebrandt_2008}
analysed the physiological properties of the bacteria that might explain their characteristic slope-type preferences.  For example,
the physical integrity of the cell membrane at different temperatures is
crucial for cell survival, and particularly its fatty acid composition
is of substantial importance.  \cite{Sikorski_Brambilla_Kroppenstedt_2008} compared the mean contents
of the fatty acids that tolerate high and low temperatures of the
\textit{B.~simplex} ecotypes.
The data showed heteroscedastic variances across putative ecotypes, and
\cite{Herberich_Sikorski_Hothorn_2010} analysed the data using
heteroscedasticity consistent covariance estimation techniques. Here,
we aim at estimating the whole conditional distribution of
the fatty acid contents (FA) of each of the six putative ecotypes
$\text{PE}_3$--$\text{PE}_7$ and $\text{PE}_9$ from $\text{GL}_1$.

The simplest conditional transformation model allowing for heteroscedasticity reads
\begin{eqnarray} \label{fattyacid_lin}
\Prob(\text{FA} \le \upsilon | \text{PE} = \text{PE}_k) =
    \Phi(\alpha_{0,k} + \alpha_k \upsilon), \quad k \in \{3, \ldots, 7, 9\}.
\end{eqnarray}
The base-learner is defined by a linear basis $\bvec_0(\upsilon) = (1,
\upsilon)^\top$ for the grid variable and a dummy-encoding basis
$\bvec_1(\text{PE}) =
(I(\text{PE}=\text{PE}_3),\ldots,I(\text{PE}=\text{PE}_9))^\top$ for the six
putative ecotypes $\{3, \ldots, 7, 9\}$.  The resulting $12$-dimensional
parameter vector $\gammavec_1$ of the \prodname product base-learner then
consists of separate intercept and slope parameters for each of the putative
ecotypes.  Since both bases are parametric, we set the penalty matrices
$\mP_0$ and $\mP_1$ to zero and choose $\lambda_0 = \lambda_1 = 0$, \ie the
base-learner does not penalise the estimates.  Note that since we assume
normality for the linear function $\alpha_{0,k} + \alpha_k \text{FA} \sim
\mathcal{N}(0, 1)$, also the fatty acid content $\text{FA}$ is assumed to be
normal with mean \textit{and} variance depending on the putative ecotype.
Since no penalisation is defined for the base-learners, penalisation of the
model parameters depends on the number of boosting iterations $M$.  For a very
large number of iterations, the algorithm converges to the estimate that is
obtained by maximising the likelihood of the linear probit model fitted to
the binary responses $\text{FA}_i \le \upsilon_\imath$, and consequently we
can fit this model by probit regression on the expanded
observations in this simple setup.

\begin{figure}[t]
\includegraphics{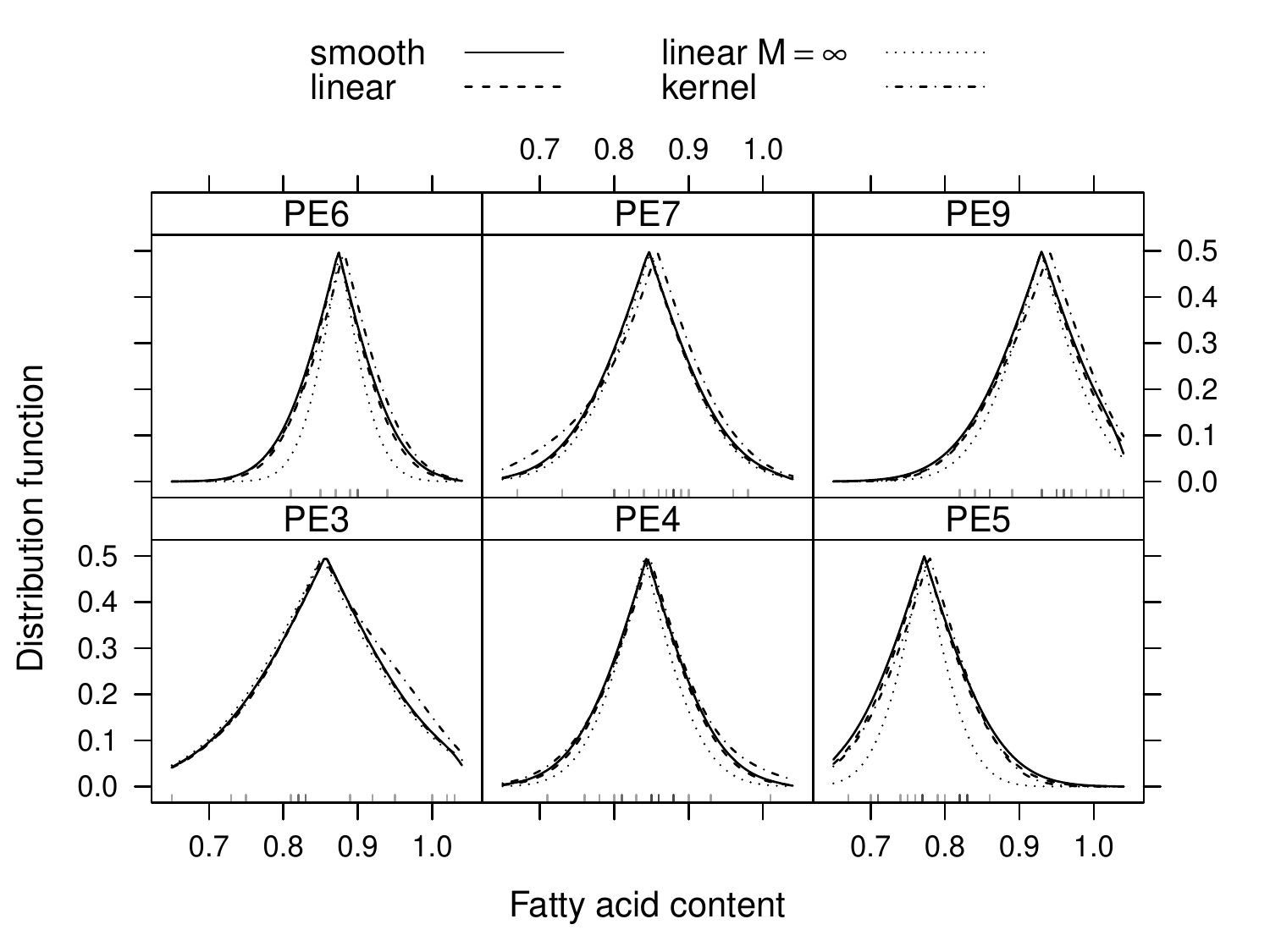}
\caption{Evolution Canyon Bacteria. Estimated conditional distribution functions
         (depicted as $pI(p < 0.5) + (1-p)I(p > 0.5)$ for probability $p$)
         of the fatty acid content of bacteria from six different putative (PE)
         ecotypes. Model (\ref{fattyacid_lin}) was fitted
         with (``linear'') and without (``linear $M = \infty$'') early stopping via the bootstrap
         whereas model (\ref{fattyacid_flex}), here denoted ``smooth'', 
         was stopped early. The observations are given as rugs. \label{fattyacid-plot}}
\end{figure}

We can relax the normal assumption on FA by allowing for more flexible
transformations in the model
\begin{eqnarray} \label{fattyacid_flex}
\Prob(\text{FA} \le \upsilon | \text{PE} = \text{PE}_k) = \Phi(h(\upsilon |
\text{PE} = \text{PE}_k)).
\end{eqnarray}
Now $\bvec_0(\upsilon)$ is a vector of $B$-spline basis functions evaluated
at $\upsilon$ for some reasonable choice of knots, while $\bvec_1$ remains as
above.  Hence, instead of assuming separate linear effects for the putative
ecotypes, we now assume separate non-parametric effects parameterised in
terms of $B$-splines.  To achieve smoothness of these non-parametric effects
along the $\upsilon$-grid, we specify the penalty matrix $\mP_0$ as
$\mP_0=\mD^\top\mD$ with second-order difference matrix $\mD$.  We do not
penalise differences between the estimated functions of different putative
ecotypes, \ie $\mP_1 = 0$, and therefore the penalty matrix of the \prodname
product is simply given by $\mP_{01}=\lambda_0\diag(\mP_0,\ldots,\mP_0)$.
Although we still assume $h(\text{FA}|\text{PE} = \text{PE}_k) \sim
\mathcal{N}(0, 1)$, the function $h$ may now be non-linear, and thus the
conditional distribution of FA given PE may be any distribution that can be
generated by a monotone transformation of a standard normal.  In this sense,
the model (\ref{fattyacid_flex}) is non-parametric.  The validity of the
normal assumption on FA is plausible when the estimated functions
$\hat{h}(\upsilon | \text{PE} = \text{PE}_k)$ are essentially linear in
$\upsilon$.  An alternative non-parametric estimate can be obtained by kernel
smoothing for mixed data types \citep{Li_Racine_2008, Hayfield_Racine_2008},
and we compare the two fits in Figure~\ref{fattyacid-plot}.

We fitted  model (\ref{fattyacid_lin}) without ($M = \infty$) and with early
stopping (via 25-fold bootstrap stratified by PE) and model
(\ref{fattyacid_flex}) with early stopping to the data and in addition
report the result of kernel smoothing, whose bandwidth was determined via
cross-validation.  The conditional distribution functions of fatty acid
content given the putative ecotype are depicted in
Figure~\ref{fattyacid-plot}.  For probabilities larger than $0.5$, the
estimated conditional distribution functions were mirrored at the $0.5$
horizontal line to allow for easier graphical inspection of
medians, variances and potential skewness.

With respect to the conditional median, the four models
lead to very similar results and support the conclusion from earlier
investigations \citep{Sikorski_Brambilla_Kroppenstedt_2008,Herberich_Sikorski_Hothorn_2010} that
the fatty acid content of \textit{B.~simplex} from putative ecotype
$\text{PE}_5$ is smaller than the others and that from $\text{PE}_9$ is larger than
the others, with the remaining ones showing no differences.
The variability cannot be assumed to be constant across different putative
ecotypes, but there is no indication of asymmetry. Early stopping and
bandwidth choice via resampling methods lead to almost the same estimated
distribution functions. Minimising the empirical risk function without early
stopping (``linear $M=\infty$''; this is equivalent to linear probit regression on
expanded observations) leads to slightly smaller variances in $\text{PE}_5$
and $\text{PE}_6$. The difficulty in discriminating between the linear model (\ref{fattyacid_lin}) and the
more flexible model (\ref{fattyacid_flex}) and the
lack of asymmetry in the plots indicate that a normal assumption on fatty
acid content is justifiable.

\subsection*{Childhood Nutrition in India}

Childhood undernutrition is one of the most urgent problems in developing and
transition countries. To provide information not only on the
nutritional status but also on health and population trends in general,
Demographic and Health Surveys (DHS) conduct nationally representative
surveys on fertility, family planning, maternal and child health, as well as
child survival, HIV/AIDS, malaria, and nutrition. The resulting data -- from
more than 200 surveys in 75 countries so far -- are available for research
purposes at \url{www.measuredhs.com}.

%\begin{figure}
%<<india-plot, echo = FALSE, fig = TRUE>>=
%
%a <- tapply(1:nrow(x), x$distH, function(i) x[i[order(x[i, "stunting"])],])
%a <- do.call("rbind", a)
%
%print(xyplot(p ~ stunting, groups = distH, data = a, type = "l",
%             col = rgb(.1, .1, .1, .1)))
%@
%\caption{India}
%\end{figure}

\setlength{\unitlength}{.9\textwidth}  % measure in textwidths

\begin{figure}
\begin{center}
  \begin{picture}(1,1)(.2, .0)
     \put(0,0){\includegraphics{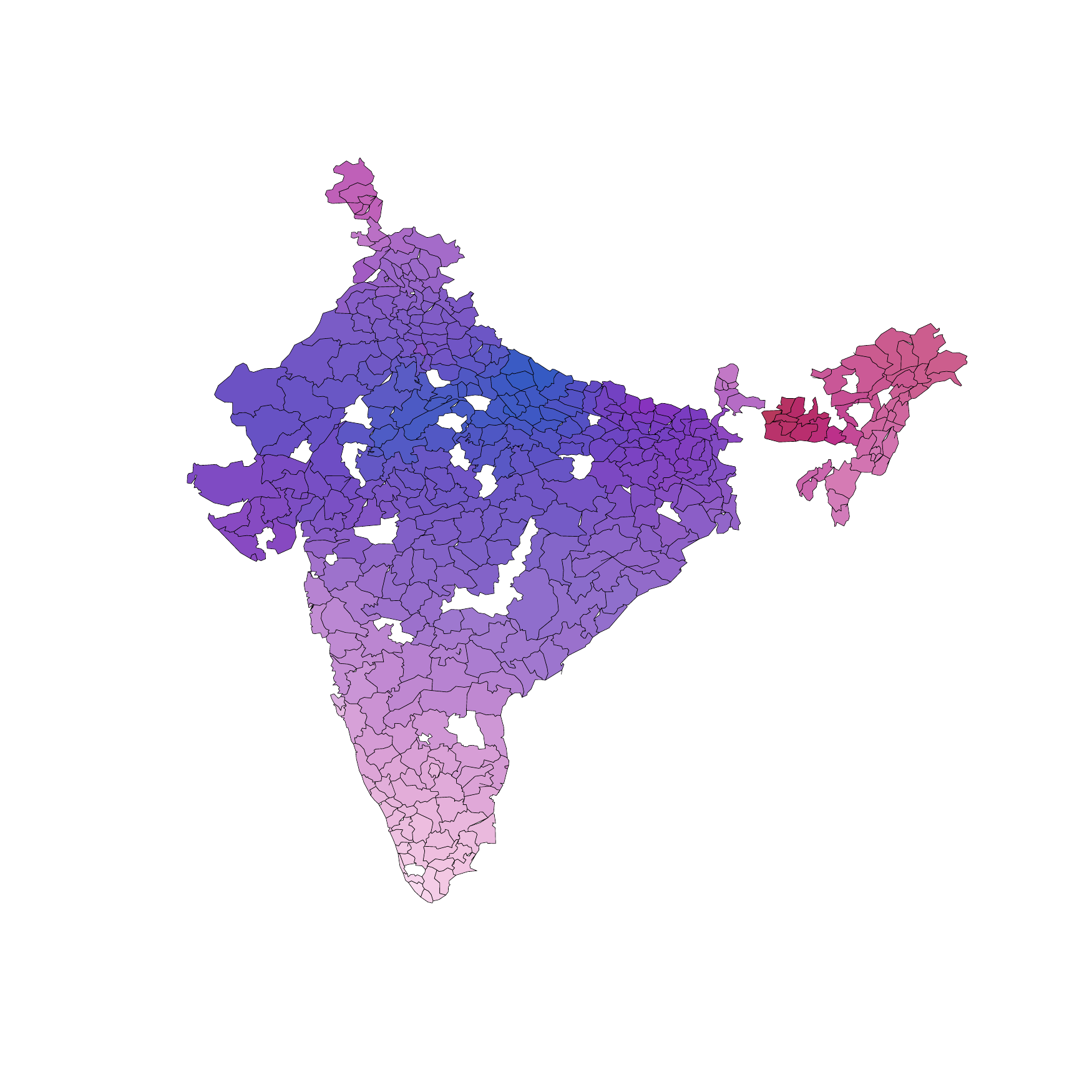}}
     \put(.7,.2){\includegraphics[width=.35\textwidth]{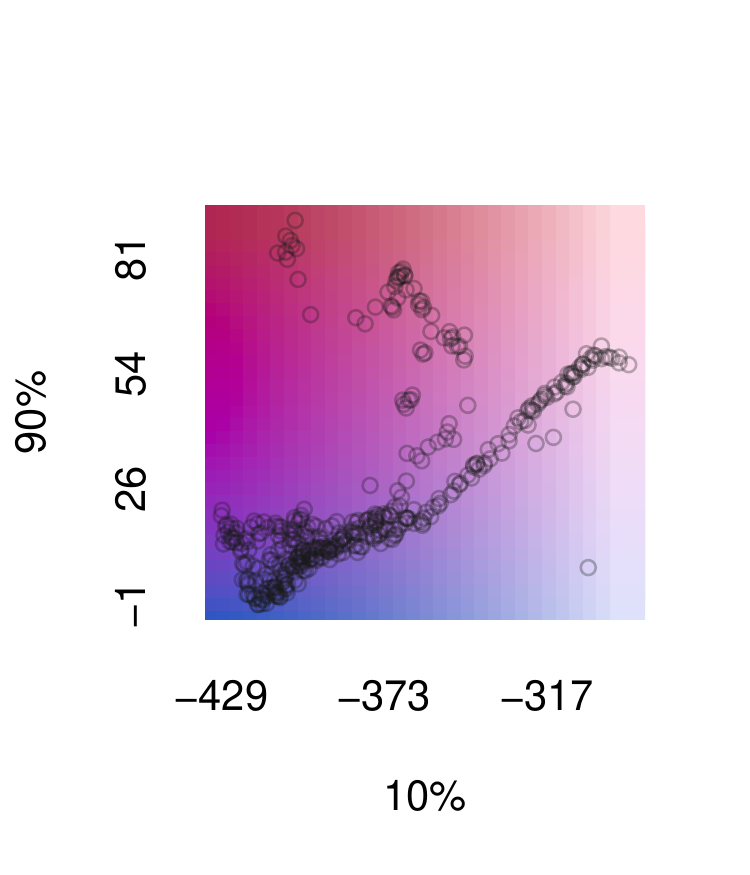}}
  \end{picture}
\end{center}
\caption{Childhood Nutrition in India. Colour-coded map of the $10\%$ and $90\%$
         conditional quantiles of the $Z$ score. Each dot in the colour legend
         corresponds to one district with the respective colour in the map.
         Blue values in the northern part
         of India correspond to small lower and upper quantiles. Red values,
         especially in the eastern Meghalaya and Assam states, indicate small
         lower quantiles but at the same time large upper quantiles. In the southern part
         of India, the lower quantiles are largest with moderate upper quantiles.
         White parts indicate districts with no observations.
         \label{india_qplot}}
\end{figure}

Childhood nutrition is usually measured in terms of a $Z$ score that compares
the nutritional status of children in the population of interest with the
nutritional status in a reference population. The nutritional
status is expressed by anthropometric characteristics, \ie height for
age; in cases of chronic childhood undernutrition, the reduced growth rate in
human development is termed stunted growth or stunting.
The $Z$ score, which compares an anthropometric characteristic of child $i$
to values from a reference population, is given as
\begin{eqnarray*}
 Z_i = \frac{\text{AC}_i-m}{s}
\end{eqnarray*}
where AC denotes the anthropometric characteristic of interest and $m$ and
$s$ correspond to median and (a robust estimate for the) standard deviation
in the reference population (stratified with respect to age, gender, and
some other variables).  While weight might be considered as the most
obvious indicator for undernutrition, we will focus on stunting, \ie
insufficient $\text{AC} = \text{height}$ for age, in the following. 
Stunting provides a measure of chronic undernutrition, whereas insufficient
weight for age might result from either acute or chronic undernutrition.  Note
that the $Z$ score, despite its name, is not assumed to be normal.

Here we focus on estimating the whole distribution of the $Z$ score measure
for childhood nutrition in India, one of the fastest growing economies and the
second-most populated country in the world.  Our investigation is based on
India's 1998--1999 Demographic and Health Survey \citep[DHS,][]{NFHS-2_2000} on
$24,166$ children visited during the survey in $412$ of the $640$ districts
of India.  The lower quantiles of this distribution can be used to assess
the severity of childhood undernutrition, whereas the upper quantiles give us
information about the nutritional status of children in families with above-average nutritional status.

The simplest conditional transformation model allowing for district-specific
means and variances reads
\begin{eqnarray*}
\Prob(Z \le \upsilon | \text{district} = k) =
    \Phi(\alpha_{0,k} + \alpha_k \upsilon), \quad k = 1, \dots, 412.
\end{eqnarray*}
The base-learner is defined by a linear basis $\bvec_0(\upsilon) = (1,
\upsilon)^\top$ for the grid variable and a dummy-encoding basis
$\bvec_1(\text{district}) =
(I(\text{district}=1),\ldots,I(\text{district}=k))^\top$ for the 412
districts.  The resulting $824$-dimensional
parameter vector $\gammavec_1$ of the \prodname product base-learner then
consists of separate intercept and slope parameters for each of the
districts of India. Note that since we assume
normality for the linear function $\alpha_{0,k} + \alpha_k Z \sim
\mathcal{N}(0, 1)$, also the $Z$ score is assumed to be
normal with mean \textit{and} variance depending on the district.

We can relax the normal assumption on $Z$ by allowing for more flexible
transformations in the model
\begin{eqnarray}
\Prob(Z \le \upsilon | \text{district} = k) =
  \Phi(h(\upsilon | \text{district} = k)), \quad k = 1, \dots, 412.
\end{eqnarray}
Now $\bvec_0(\upsilon)$ is a vector of $B$-spline basis functions evaluated
at $\upsilon$ for some reasonable choice of knots, while $\bvec_1$ remains as
above.  Hence, instead of assuming separate linear effects for the
districts, we now assume separate non-parametric effects parameterised in
terms of $B$-splines.  To achieve smoothness of these non-parametric effects
along the $\upsilon$-grid, we specify the penalty matrix $\mP_0$ as
$\mP_0=\mD^\top\mD$ with second-order difference matrix $\mD$.  It 
makes sense to induce spatial smoothness on the conditional distribution
functions of neighbouring districts since we do not expect the distribution
of the $Z$ score to change much from one district to its neighbouring
districts.  In fact, spatial smoothing is absolutely necessary in this example
since otherwise we would estimate $412$ separate distribution functions for
the districts in India.  To implement spatial smoothness of neighbouring
districts, the penalty matrix $\mP_1$ is chosen as an adjacency matrix, where
the off-diagonal elements indicate whether two districts are neighbours
(represented with a value of $-1$) or not (represented with a value of $0$).
The diagonal of the adjacency matrix contains the number of neighbours for
the corresponding district.  The estimated conditional transformation
function $\hat{h}(Z | \text{district} = k)$ can be interpreted as a
transformation of the $Z$ scores in district $k$ to standard normality.
Because the number of observations is large and the base-learner is fitted
with penalisation, we stopped the boosting algorithm according to the
in-sample empirical risk.

From the estimated conditional distribution functions, we compute
the $\tau$ quantiles of the $Z$ score for each district via
\begin{eqnarray*}
\hat{Q}(\tau | \text{district} = k) = \text{inf}\{\upsilon :
  \Phi(\hat{h}(\upsilon | \text{district} = k) \ge \tau\}.
\end{eqnarray*}
The conditional $10\%$ and $90\%$ quantiles are depicted in a colour-coded
map in Figure~\ref{india_qplot}. The spatially smooth estimated
lower and upper conditional quantiles shown simultaneously allow 
differentiation between three groups of districts: (A) districts with
small lower and upper conditional quantiles (blue, especially in the
Uttar Pradesh state), where the $Z$ score is stochastically smaller than
that of the remaining parts of India and thus all children are less well fed;
(B) districts with more severe inequality, \ie
small lower but at the same time large upper quantiles (red, in the
Meghalaya and Assam states); and (C) districts with relatively large
lower and upper quantiles, which indicates a relatively good nutrition status
of all children in the southern districts of India (violet, in Andhra Pradesh,
Madhya Pradesh, Maharashtra, Tamil Nadu, and Kerala).

\subsection*{Head Circumference Growth}

The Fourth Dutch Growth Study \citep{Fredriks_Buuren_Burgmeijer_2000}
is a cross-sectional study that measures growth and development of the
Dutch population between the ages of 0 and 22 years. The study measured, among
other variables, head circumference (HC) and age of 7482 males and 7018
females. \cite{Stasinopoulos_Rigby_2007} analysed the
head circumference of $7040$ males with explanatory variable age using
a GAMLSS model with a Box-Cox $t$ distribution describing the first four
moments of head circumference conditionally on age. The models show
evidence of kurtosis, especially for older boys.
We estimate the whole conditional distribution function via
the conditional transformation model
\begin{eqnarray*}
\Prob(\text{HC} \le \upsilon | \text{age} = x) =
  \Phi(h(\upsilon | \text{age} = x)).
\end{eqnarray*}
The base-learner is the \prodname product of $B$-spline basis
functions $\bvec_0(\upsilon)$ for head circumference and
$B$-spline basis functions for $\text{age}^{1/3}$. The root transformation
just helps to cover the data better with equidistant knots. The penalty
matrices $\mP_0$ and $\mP_1$ penalise second-order differences, and thus
$\hat{h}$ will be a smooth bivariate \prodname product spline of head
circumference and age. It is important to note that smoothing takes
place in both dimensions. Consequently, the conditional distribution
functions will change only slowly with age, which is a reasonable
assumption. Since the number of observations is also large, we
stopped the algorithm based on the in-sample empirical risk.

\begin{figure}[t]
\includegraphics{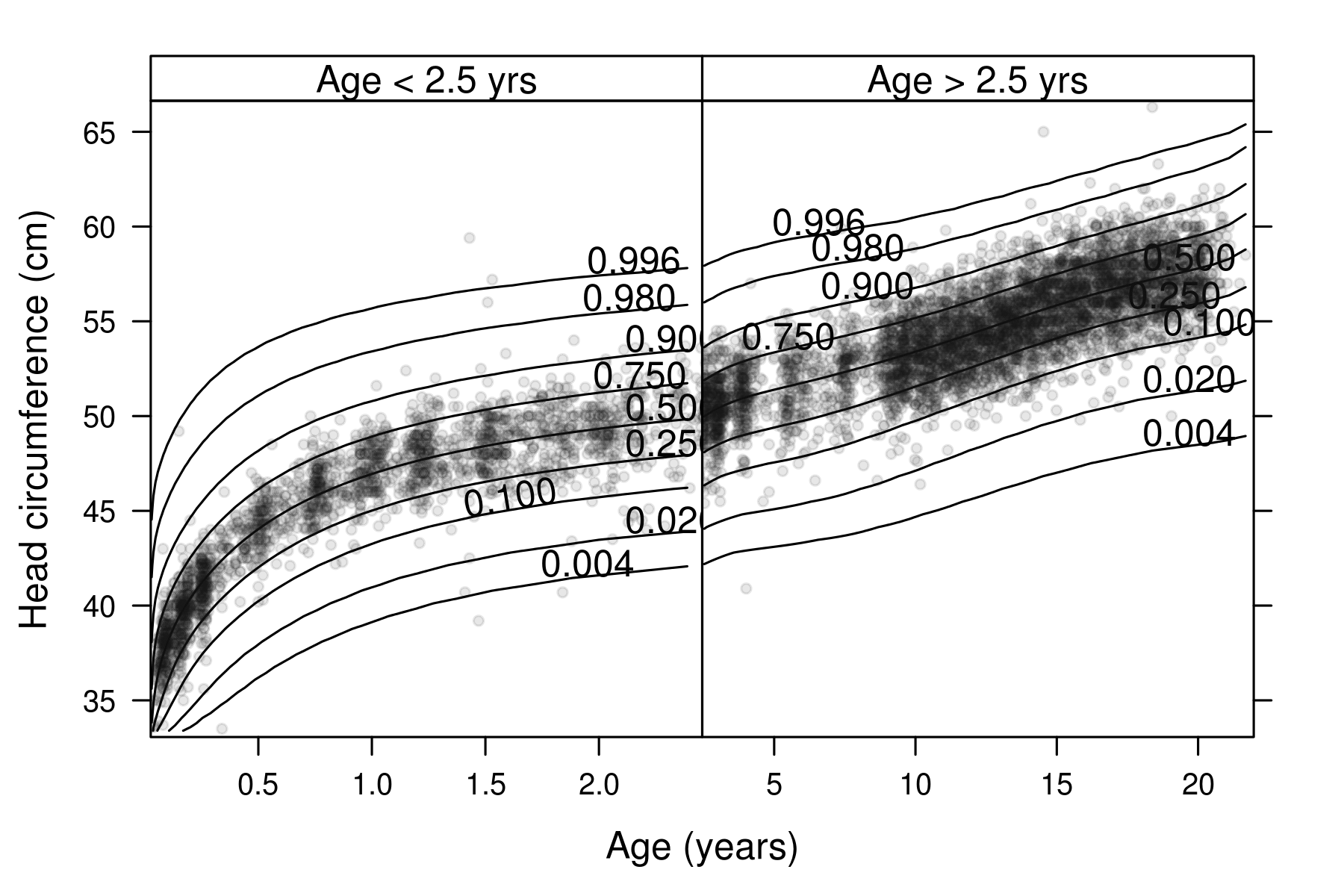}
\caption{Head Circumference Growth. Observed head circumference and age for 
         $7040$ boys with estimated quantile curves for 
         $\tau = 0.04, 0.02, 0.1, 0.25, 0.5, 0.75, 0.9, 0.98, 0.996$.
         \label{heads-plot}}
\end{figure}

Figure~\ref{heads-plot} shows the data overlaid with quantile
curves obtained via inversion of the estimated conditional distributions.
The figure can be directly compared with Figure~16 of
\cite{Stasinopoulos_Rigby_2007} and also indicates a certain asymmetry
towards older boys.

%%, pdf = FALSE, png = TRUE}

\subsection*{Deer-vehicle Collisions}

Collisions of vehicles with roe deer are a serious threat to
human health and animal welfare. In Bavaria, Germany, more than
$40,000$ deer-vehicle collisions (DVCs) take place every year.
\cite{Hothorn_Brandl_Mueller_2011} investigated the spatial distribution
of the risk of deer-vehicle collisions; here we focus on the temporal
aspect of the risk for two years, 2006 and 2009. For all $74,650$
collisions reported to the police in these two years, we attributed
each accident to the specific day of the year.

Although the number of DVCs is a discrete random variable, the distribution
of the number of DVCs conditional on the day of the year can be estimated
by means of an appropriate base-learner using the model
\begin{eqnarray*}
\Prob(\text{DVCs} \le \upsilon | \text{day} = x_1, \text{year} = x_2) =
  \Phi(h_1(\upsilon | \text{day} = x_1) + h_2(\upsilon | \text{day} = x_1, \text{year} =
x_2)).
\end{eqnarray*}
Here, $\hat{\mu}$ is the counting measure with support $\upsilon_1, \dots,
\upsilon_N$ equal to the support of the empirical distribution of the
response.  Conceptually, the basis function $\bvec_0$ should allow for $n =
N$ parameters (one for each $\upsilon_\imath$), whose first-order differences
should not become too large.  To restrict the number of parameters
in the base-learners, we use $B$-splines to approximate such a discrete
function on the $\upsilon$-grid.  It should further be noted that the day of
year is a discrete cyclic random variable.  Therefore, we chose
$\bvec_1(x_1)$ as cyclic $B$-splines of the day, which are obtained by a simple
modification of the $B$-spline design matrix and the difference penalty that
results from fusing the two ends of the co-domain.  In analogy, a cyclic
$B$-spline is applied to the varying coefficient term $\bvec_2(x_1, x_2) =
\bvec_1(x_1) \times I(x_2 = 2009)$, which captures temporal differences
between the two years and yields a cyclic $B$-spline of the days in 2009.
Since the data are discrete, we only penalise first-order differences in
both base-learners.

\begin{figure}[t]
\includegraphics{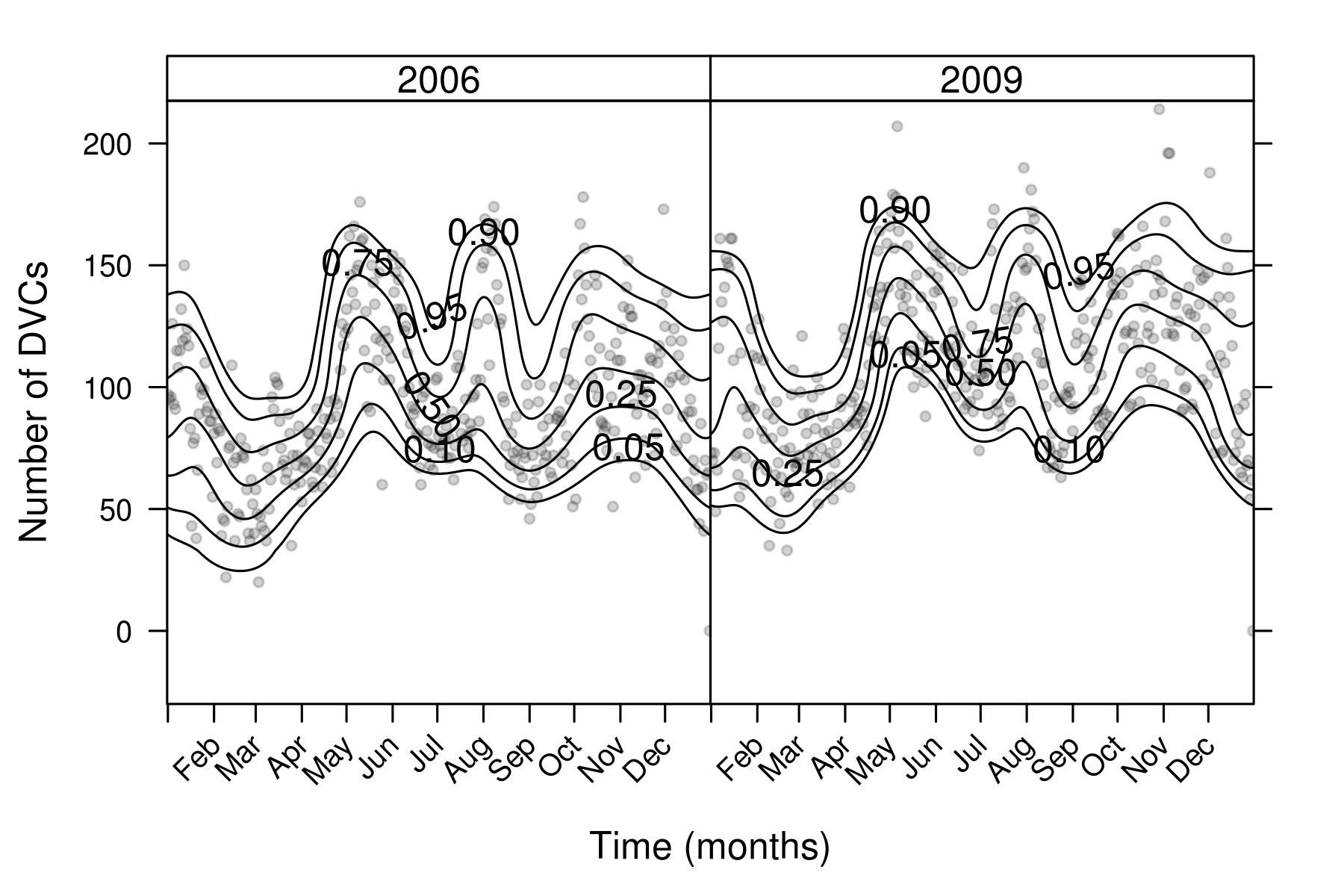}
\caption{Deer-vehicle Collisions. Number of deer-vehicle collisions (DVCs) per day in
         2006 and 2009 in Bavaria, Germany, with estimated quantile curves for 
         $\tau = 0.05, 0.1, 0.25, 0.5, 0.75, 0.9, 0.95$. \label{DVC-plot}}
\end{figure}

Figure~\ref{DVC-plot} shows three risk peaks. The first one occurs early in
May -- the beginning of the growing and buck hunting season -- and ends mid-June.
A second and sharper peak is observed in the first week of August and
corresponds to the mating season of roe deer. After a low-risk period of
approximately six weeks, the risk starts to increase again at the beginning
of October and slowly decreases until April for reasons yet unknown. Note
that the distribution in 2009 has a larger median than that in 2006 but also
shows less extreme peaks.

\subsection*{Birth Weight Prediction}

Recent advances in neonatal medicine have lowered the threshold of survival
to a gestational age of 23--24 weeks and to a birth weight of approximately
500 g.  As neonatal risks of morbidity and mortality are highest in the
lowest weight range, diagnostic assessment of the small foetus needs to be
as precise as possible.  \cite{Schild_Maringa_Siemer_2008} focused on this
high-risk group of small foetuses ($\le 1600$ g) and proposed a formula for
estimating birth weight based on ultrasound imaging performed within seven
days before delivery.  In addition to predicting the expected birth weight
given four standard 2D ultrasound parameters (HC: head circumference, FE: femur length, BPD: biparietal
diameter, and AC: transverse diameter and circumference
of the foetal abdomen) and three additional 3D ultrasound parameters (UA: upper arm volume, FEM: thigh volume, and
ABDO: abdominal volume) $\mX \in \RR^7$, we aim at assessing the uncertainty
of this prediction.  The data on 150 predominantly Caucasian women,
collected in a prospective cohort study
at the universities in Bonn and Erlangen, Germany, analysed by
\cite{Schild_Maringa_Siemer_2008}, were utilised to derive $80\%$
prediction intervals for birth weight (BW).

We begin with the linear model estimated by \cite{Schild_Maringa_Siemer_2008}
\begin{eqnarray*}
\text{BW}_\xvec & = & 656.41 + 1.832 \times \text{ABDO} +
                               31.198 \times \text{HC} +
                               5.779 \times \text{FEM} + \\
                           & & 73.521 \times \text{FL} +
                               8.301 \times \text{AC} -
                               449.886 \times \text{BPD} +
                               32.534 \times \text{BPD}^2 + \\
                           & & 77.465 \times \Phi^{-1}(U),
\end{eqnarray*}
and the classical prediction interval for a foetus with
ultrasound parameters $\xvec$ is then the symmetric interval around
the estimated conditional mean $\hat{\Ex}(\text{BW} | \mX = \xvec)$,
whose width is given by $2 \times t_{150 - 8, 0.9} \times 77.465 \times \sqrt{1 +
\Var(\hat{\Ex}(\text{BW} | \mX = \xvec))}$.

The normal assumption can be relaxed by deriving the upper and lower
conditional quantiles from two quantile regression models.
Linear quantile regression \citep{Koenker_Bassett_1978}
for the conditional $10\%$, $50\%$ and $90\%$ quantiles assumes that
\begin{eqnarray*}
\text{BW}_\xvec = \alpha_{0, \tau} + \xvec^\top \alphavec_\tau + Q_\tau(U),
                  \quad \text{for } \tau = 0.1, \tau = 0.5, \text{ and } \tau = 0.9
\end{eqnarray*}
with $Q_\tau(\tau) = 0$. The corresponding prediction interval
for a foetus with ultrasound parameters $\xvec$ is now $(\hat{\alpha}_{0, 0.1} + \xvec^\top
\hat{\alphavec}_{0.1}, \hat{\alpha}_{0, 0.9} + \xvec^\top \hat{\alphavec}_{0.9})$.
A more flexible description of the functional relationship between
ultrasound parameters and quantiles is given by the additive quantile regression
model \citep{Koenker_Ng_Portnoy_1994}
\begin{eqnarray*}
\text{BW}_\xvec = \alpha_{0, \tau} + \sum_{j = 1}^7 r_{j,\tau}(x_j) +
Q_\tau(U), \quad \text{for } \tau = 0.1, \tau = 0.5, \text{ and } \tau = 0.9.
\end{eqnarray*}
Here, $r_{j, \tau}$ is a quantile-specific smooth function of the
$j$th ultrasound parameter. Parameter tuning is difficult
for these models; we therefore applied a boosting approach to additive
quantile regression
\citep[with early stopping via 25-fold bootstrap]{Fenske_Kneib_Hothorn_2011}.
Prediction intervals can now be derived by
$(\hat{\alpha}_{0, 0.1} + \sum_{j = 1}^7
\hat{r}_{j,0.1}(x_j), \hat{\alpha}_{0, 0.9} + \sum_{j = 1}^7
\hat{r}_{j,0.9}(x_j))$. Note that for either quantile regression model,
the prediction interval is based on two separate models: one for $\tau =
0.1$ and one for $\tau = 0.9$.

Finally, we derive prediction intervals from the conditional transformation
model
\begin{eqnarray*}
\Prob(\text{BW} \le \upsilon | \mX = \xvec) =
  \Phi\left(\sum_{j = 1}^7 h_j(\upsilon | x_j)\right)
\end{eqnarray*}
where, under the assumption of additivity of the transformation function $h$,
each ultrasound parameter may influence the moments of the conditional birth
weight distribution.  The $j$th base-learner is the \prodname product of
$B$-spline basis functions $\bvec_0(\upsilon)$ for birth weight and
$\bvec_j(x_j)$ are $B$-spline basis functions for the $j$th ultrasound
parameter.  The penalty matrices $\mP_0$ and $\mP_j$ penalise second-order
differences, and thus all estimates $\hat{h}_j$ will be smooth bivariate
\prodname product splines of birth weight and the respective ultrasound
parameter, with both dimensions being subject to smoothing.  The number of
boosting iterations was determined by 25-fold bootstrap.  From the estimated
conditional distribution functions, we compute the $\tau$ quantiles of the
birth weight via
\begin{eqnarray*}
\hat{Q}(\tau | \mX = \xvec) = \text{inf}\left\{\upsilon :
  \Phi\left(\sum_{j = 1}^7 \hat{h}_j(\upsilon | x_j)\right) \ge
\tau\right\}
\end{eqnarray*}
and derive the prediction interval as $(\hat{Q}(0.1 | \mX = \xvec), \hat{Q}(0.9 | \mX =
\xvec))$. Note that, unlike in the quantile regression approach, the prediction
interval obtained from the conditional transformation model
is based on only one model that describes the whole conditional
distribution of birth weight.

\begin{figure}
\includegraphics{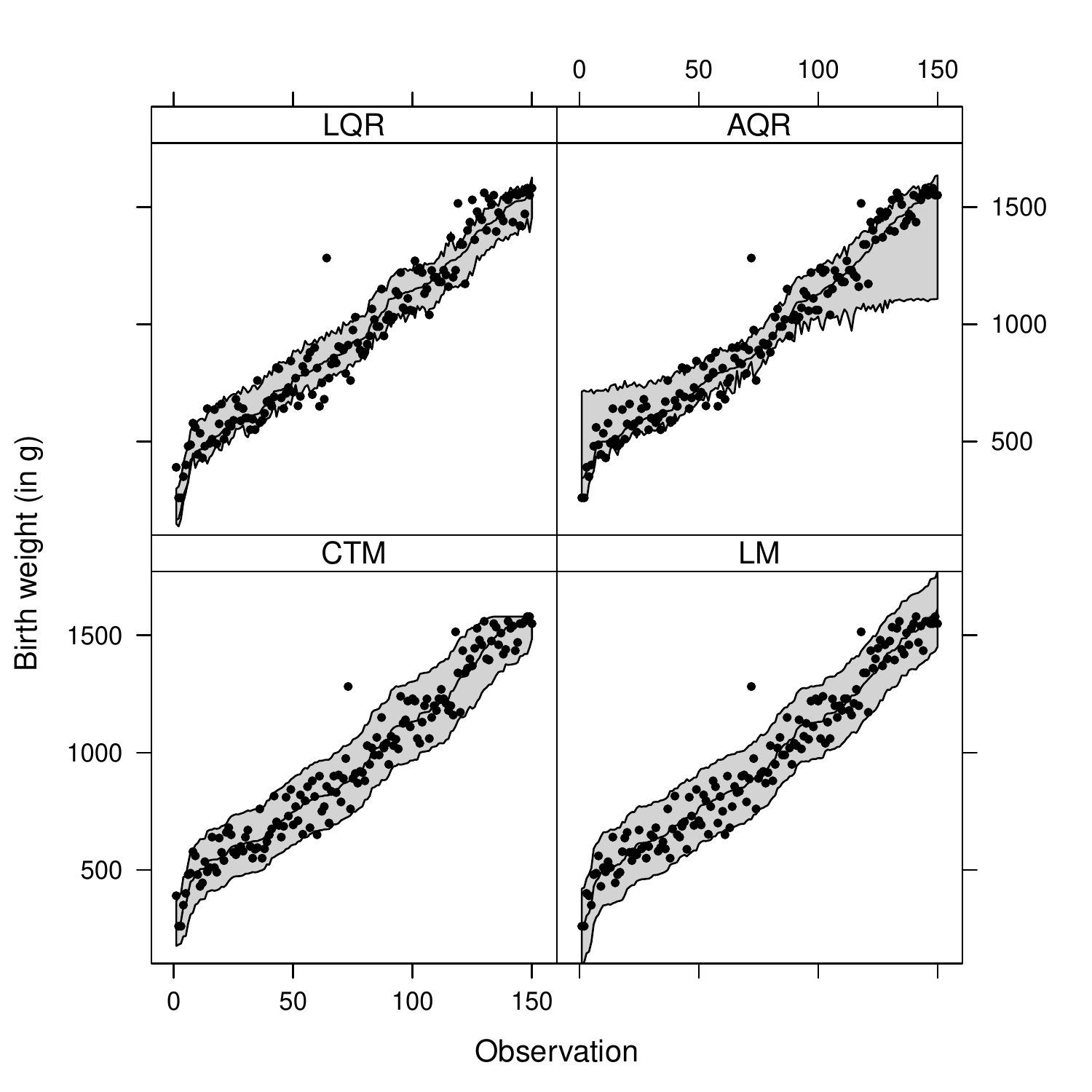}
\caption{Birth Weight Prediction. Observed birth weights for 150 small foetuses (dots), ordered with
         respect to the estimated mean or median expected birth weight (central black line). The 
         shaded area represents foetus specific $80\%$ prediction intervals for the
         linear model (LM), linear quantile regression model (LQR), additive quantile 
         regression model (AQR) and conditional transformation model (CTM). \label{fetus-plot}}
\end{figure}

The observed birth weights ordered with respect to the predicted mean
(linear model) or median (quantile regression and conditional transformation
model) are depicted in Figure~\ref{fetus-plot}.  In addition, the respective
$80\%$ prediction intervals are visualised by grey areas.  It must be noted
that, for all models, the prediction intervals are only interpretable for
future observations; however, poor coverage for the learning sample also
indicates poor coverage for future cases.  The prediction intervals obtained
from linear quantile regression indicate that the model is confident
about its predictions over the whole range of birth weights.  This is also
the case for the additive quantile regression models for birth weights of
approximately 1000 g, but the uncertainty increases for very small and larger
foetuses.  The intervals obtained from the linear model and the conditional
transformation model appear to be similar.  For birth weights between 500 and
1400 g, the prediction intervals of the conditional transformation model are
symmetric around the median.  This might be an indication that the
normal assumption by the linear model is not completely unrealistic.  The
smaller interval widths that can be seen for the linear model are most
likely due to the variance estimate in this case ignoring the
model choice process that was performed prior to the final model fit by
\cite{Schild_Maringa_Siemer_2008}.  The conditional transformation model
takes this variability into account.  The results may also be an indication
that the assumption of additivity of the transformation function 
rather than of the regression function (for quantile regression             
models) might be more appropriate for modelling birth weights.

%% \begin{comment}

\subsection*{Beyond Mean Boston Housing Values}

The Boston Housing data, first published by \cite{Harrison_Rubinfeld_1978}
and later corrected and spatially aligned by \cite{Gilley_Pace_1996}, have
become a standard test-bed for variable selection and model choice.  Almost
exclusively, the 13 explanatory variables have been selected with respect to
their influence on the mean or median of the conditional median house value
in a certain tract.  Assuming a conditional transformation model, we attempt to
detect dependencies of higher moments of the conditional median house value
from the explanatory variables.  We focus on the 12 numeric explanatory
variables and ignore the binary variable coding for Charles River boundary
in the conditional transformation model
\begin{eqnarray*}
\Prob(\text{MEDV} \le \upsilon | \mX = \xvec) & = &
  \Phi\left(\alpha_\text{tract} + h_0(\upsilon | 1) +
            \sum_{j = 1}^{12} h_{j}(1 | x_j) +
            \sum_{j = 1}^{12} h_{j}(\upsilon | x_j)\right).
\end{eqnarray*}
In this model, $\alpha_\text{tract}$ is a tract-specific, spatial random
effect, whose correlation structure is determined by a Markov random field
defined by the neighbouring structure of the tracts capturing spatial
autocorrelation and heterogeneity (similar as in the example on childhood
nutrition in India).  The term $h_0(\upsilon | 1)$ is an unconditional
transformation of the median house value, \ie this transformation is
independent of the explanatory variables.  The explanatory variables may
influence the mean of the transformed median house value $h_0(\text{MEDV} |
1)$ via $h_\xvec(\xvec) = \sum_{j = 1}^{12} h_j(1 | x_j)$ only or may also
affect higher moments via the interaction terms $\sum_{j = 1}^{12}
h_{j}(\upsilon | x_j)$.  The latter term extends the transformation model
$h_0(\text{MEDV} | 1) + \sum_{j = 1}^{12} h_j(1 | x_j)$ to a conditional
transformation model.  The base-learners for the transformation function
$h_0(\upsilon | 1)$, the effects $h_j(1|x_j)$ and the interaction
terms $h_{j}(\upsilon | x_j)$ are constructed based on cubic $B$-spline basis
functions supplemented with second-order difference penalty.  More
specifically, $\bvec_j(\xvec)$ and $\bvec_0(\upsilon)$ are both represented
in terms of a reparameterisation of the $B$-spline basis functions that allows
separation of the non-linear terms into a constant, a linear effect and the
non-linear (orthogonal) deviation from the linear effect, \ie
\begin{eqnarray*}
 \bvec_j(\xvec) = 1 + x_j + \tilde{\bvec}_j(x_j)\quad\mbox{and}\quad 
 \bvec_0(\upsilon)=1+\upsilon+\tilde{\bvec}_0(\upsilon),
\end{eqnarray*}
where $\tilde{\bvec}_j(x_j)$ and $\tilde{\bvec}_0(\upsilon)$ are the
non-linear deviation effects.  Taking the \prodname product after applying
the decomposition yields a decomposition into linear and non-linear main
effects of $x_j$ and $\upsilon$ as well as linear and non-linear
interaction terms \citep[see][for technical details of this
decomposition]{Fahrmeir_Kneib_Lang_2004, Kneib_Hothorn_Tutz_2009}.  The
advantage of this expanded parameterisation is that the automatic model
choice capabilities of the boosting algorithm allow us to flexibly determine
whether linear or non-linear effects are required and whether there actually
is an interaction between the transformation function and specific
effects of explanatory variables.

\begin{figure}
\includegraphics[width = \textwidth]{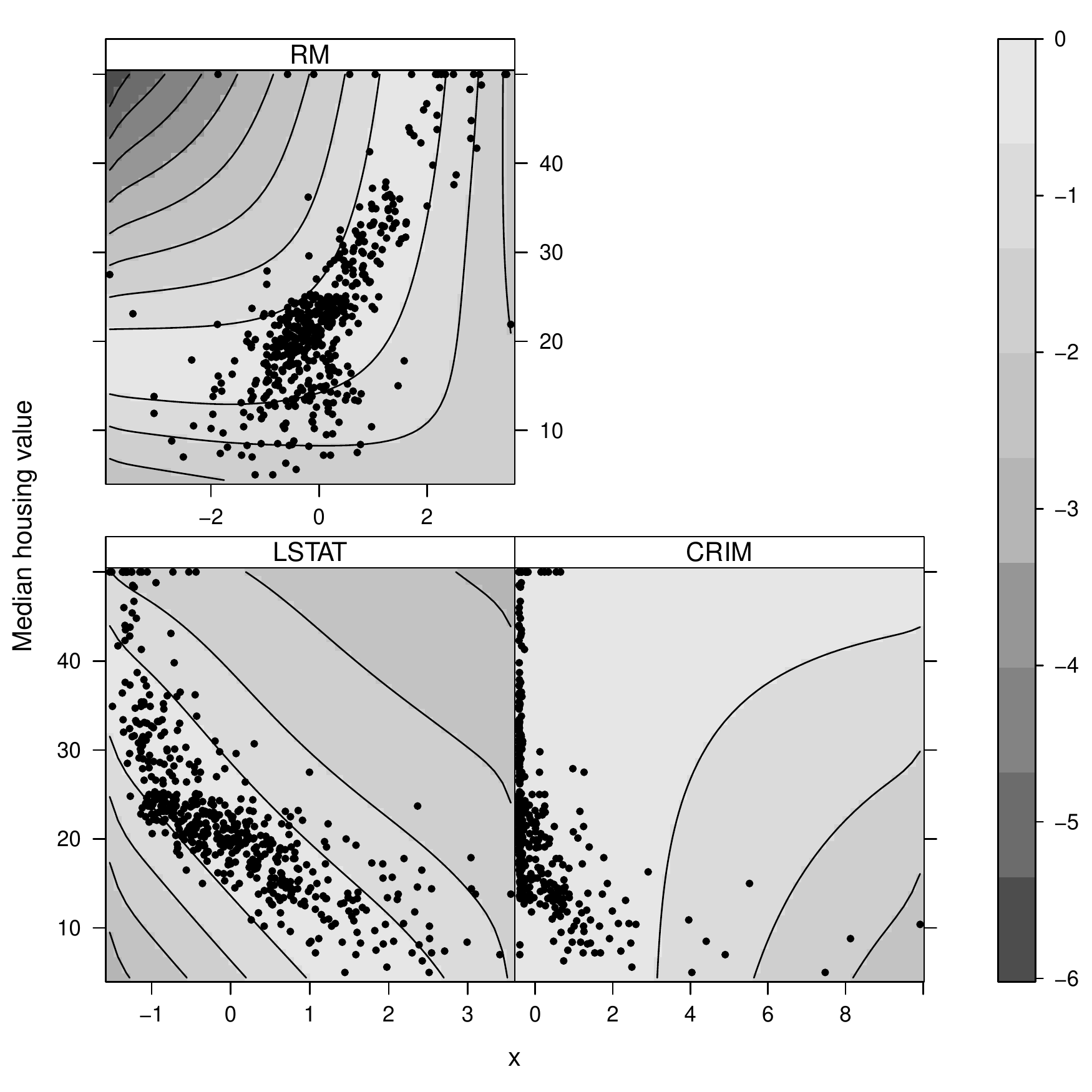}
%<<BostonHousing-plotp, echo = FALSE, fig = TRUE>>=
%print(plot(mod))
%@
\caption{Beyond Mean Boston Housing Values. Conditional transformation model
         for the three selected variables per capita crime (CRIM),
         average numbers of rooms per dwelling (RM) and percentage values of lower status
         population (LSTAT). Each panel depicts the data as scatter plots
         along with the corresponding negative absolute values of the  estimated transformation function
         at the probit scale. The explanatory variables were standardised 
         prior to analysis. \label{BostonHousing-plot}}
\end{figure}

Censored observations were dealt with by choosing inverse probability of
censoring weights $w_i$ for the empirical risk function (\ref{mod:risk})
derived from the Kaplan-Meier estimate of the censoring distribution.
The stability selection procedure \citep{Meinshausen_Buehlmann_2010}
selected three variables that have an influence on the conditional
distribution of the median housing value (MEDV), namely per capita crime (CRIM),
average numbers of rooms per dwelling (RM), and percentage values of lower status
population (LSTAT). After variable selection, we refitted a conditional
transformation model of the simpler form
\begin{eqnarray*}
& & \Prob(\text{MEDV} \le \upsilon | \text{CRIM}, \text{RM},\text{LSTAT}) \\
& & =   \Phi\left(h_\text{CRIM}(\upsilon | \text{CRIM}) +
            h_\text{RM}(\upsilon | \text{RM})  +
            h_\text{LSTAT}(\upsilon | \text{LSTAT})\right),
\end{eqnarray*}
where the base-learners are \prodname products of $B$-spline bases. The fitted
functions can be conveniently depicted in the observation space.  For
example, a scatter plot of MEDV and CRIM and a grey-level image of the
bivariate function $\hat{h}_\text{CRIM}(\text{MEDV} | \text{CRIM})$ can be
viewed in the same coordinate system.  Similar to the mirrored distribution
functions in Figure~\ref{fattyacid-plot}, we show negative absolute values
of the fitted functions $\hat{h}$ for easier interpretation.

Figure~\ref{BostonHousing-plot} indicates that
the percentage values of lower status population (LSTAT) lead to
smaller values of the median housing value at almost constant variance.
However, the conditional distribution will be skewed towards higher MEDV
values. For tracts with small average numbers of rooms per dwelling (RM),
the median housing value is small and increases with increasing numbers of
rooms. The same applies to the variability, since the estimated function
$\hat{h}_\text{RM}(\text{MEDV} | \text{RM})$ shows more spread for larger
values of RM.  Per capita crime seems to have an effect on variability and
skewness, since for larger crime values, the distribution will be heavily
skewed and less variable than small per capita crime values. However,
compared to the other two variables, the influence is only of marginal
value due to small absolute contributions of this model term to the full
model.

%% \end{comment}

\section{Empirical Evaluation} \label{sec:eval}

We shall compare the empirical performance of conditional transformation
models fitted by means of the proposed boosting algorithm to two
competitors.  Conditional transformation models are semiparametric models in
the sense that we assume a certain distribution for the transformed
responses and additivity of the model terms on the scale of the
corresponding quantile function.  Therefore, it is natural to compare
the goodness of the estimated conditional distribution functions to a
fully parametric approach and a non-parametric estimation technique.

For the sake of simplicity, we study a model in which two explanatory variables
influence both the conditional expectation and the conditional variance of
a normal distributed response $Y$. The error term $\Phi^{-1}(U)$ is standard normal,
and, to obtain normal responses, we restrict the possible transformations
to linear functions:
\begin{eqnarray*}
\Phi^{-1}(U) & = & h(\Yx |\xvec) 
      = \sum_j h_j(\Yx | \xvec) 
      = \sum_j b_j(\xvec)\Yx - a_j(\xvec) \\
& = & \Yx \sum_j b_j(\xvec) - \sum_j a_j(\xvec) \\
\iff \Yx & = & \frac{\Phi^{-1}(U) + \sum_j a_j(\xvec)}{\sum_j b_j(\xvec)} 
      \sim \calN\left(\frac{\sum_j a_j(\xvec)}{\sum_j b_j(\xvec)}, 
                      \left(\sum_j b_j(\xvec)\right)^{-2}\right).
\end{eqnarray*}
Although the partial transformation functions are linear in
$\Yx$, the expectation and variance
depend on the explanatory variables in a non-linear way. The choices
$X_1 \sim \calU[0, 1], X_2 \sim \calU[-2, 2]$, 
$a_1(\xvec) = 0, a_2(\xvec) = x_2$, and 
$b_1(\xvec) = x_1, b_2(\xvec) = 0.5$ lead to the heteroscedastic 
varying coefficient model
\begin{eqnarray} \label{simmod}
\Yx = \frac{1}{x_1 + 0.5}x_2 + \frac{1}{x_1 + 0.5}\Phi^{-1}(U),
\end{eqnarray}
where the variance of $\Yx$ ranges between $0.44$ and $4$ depending on
$X_1$. This model can be fitted in the GAMLSS framework under the
assumptions that the expectation of the normal response depends on a 
smoothly varying regression coefficient $(X_1 + 0.5)^{-1}$ for $X_2$ and
that the variance is a smooth function of $X_1$. This model is therefore fully
parametric. As a non-parametric counterpart, we use a kernel estimator for
estimating the conditional distribution function of $\Yx$ as a function
of the two explanatory variables.

The conditional transformation model
\begin{eqnarray*}
\Prob(Y \le \upsilon | X_1 = x_1, X_2 = x_2) = \Phi(h(\upsilon| x_1, x_2)) = 
    \Phi(h_1(\upsilon | x_1) + h_2(\upsilon | x_2))
\end{eqnarray*}
is a semiparametric compromise between these two extremes. 
The error distribution is assumed to be standard normal and additivity
of the transformation function $h$ is also part of the model specification.
The base-learners are \prodname products of $B$-spline basis
functions $\bvec_0(\upsilon)$ for $Y$ and
$B$-spline basis functions for $X_1$ and $X_2$, respectively.  The penalty
matrices $\mP_0$, $\mP_1$ and $\mP_2$ penalise second-order differences, and thus
$\hat{h}_j$ will be smooth bivariate \prodname product splines of 
the response and explanatory variables $X_1$ and $X_2$. Smoothing takes
place in both dimensions. 

For all three approaches, we obtain estimates of 
$\Prob(Y \le \upsilon | X_1 = x_1, X_2 = x_2)$ over a grid on 
$x_1, x_2$ and compute the mean absolute deviation (MAD)
of the true and estimated probabilities
\begin{eqnarray*}
\text{MAD}(x_1, x_2) = \frac{1}{n} \sum_{\upsilon} 
    |\Prob(Y \le \upsilon | X_1 = x_1, X_2 = x_2) - 
     \hat{\Prob}(Y \le \upsilon | X_1 = x_1, X_2 = x_2)|
\end{eqnarray*}
for each pair of $x_1$ and $x_2$. Then, the minimum, the median, and the maximum
of the MAD values over this grid are computed as summary statistics.
This procedure was repeated for $100$ random samples of size $N = 200$ 
drawn from model (\ref{simmod}). Cross-validation was used to determine the
bandwidths for the kernel-based methods, for details see
\cite{Hayfield_Racine_2008}. The boosting-based estimation of 
GAMLSS models \citep{Mayr_Fenske_Hofner_2012} turned out to be more stable 
than the reference implementation \citep[package \pkg{gamlss},][]{pkg:gamlss}, and
we therefore fitted the GAMLSS models by the dedicated boosting algorithm. 
For GAMLSS and conditional transformation models fitted by boosting,
the number of boosting iterations was determined via sample-splitting.
To investigate the stability of the three procedures under non-informative
explanatory variables, we added $p = 1, \dots, 5$ uniformly distributed variables
without association to the response to the data and included them as potential
explanatory variables in the three models. The case $p = 0$ corresponds to model~(\ref{simmod}).

\begin{sidewaysfigure}
\begin{center}
\includegraphics{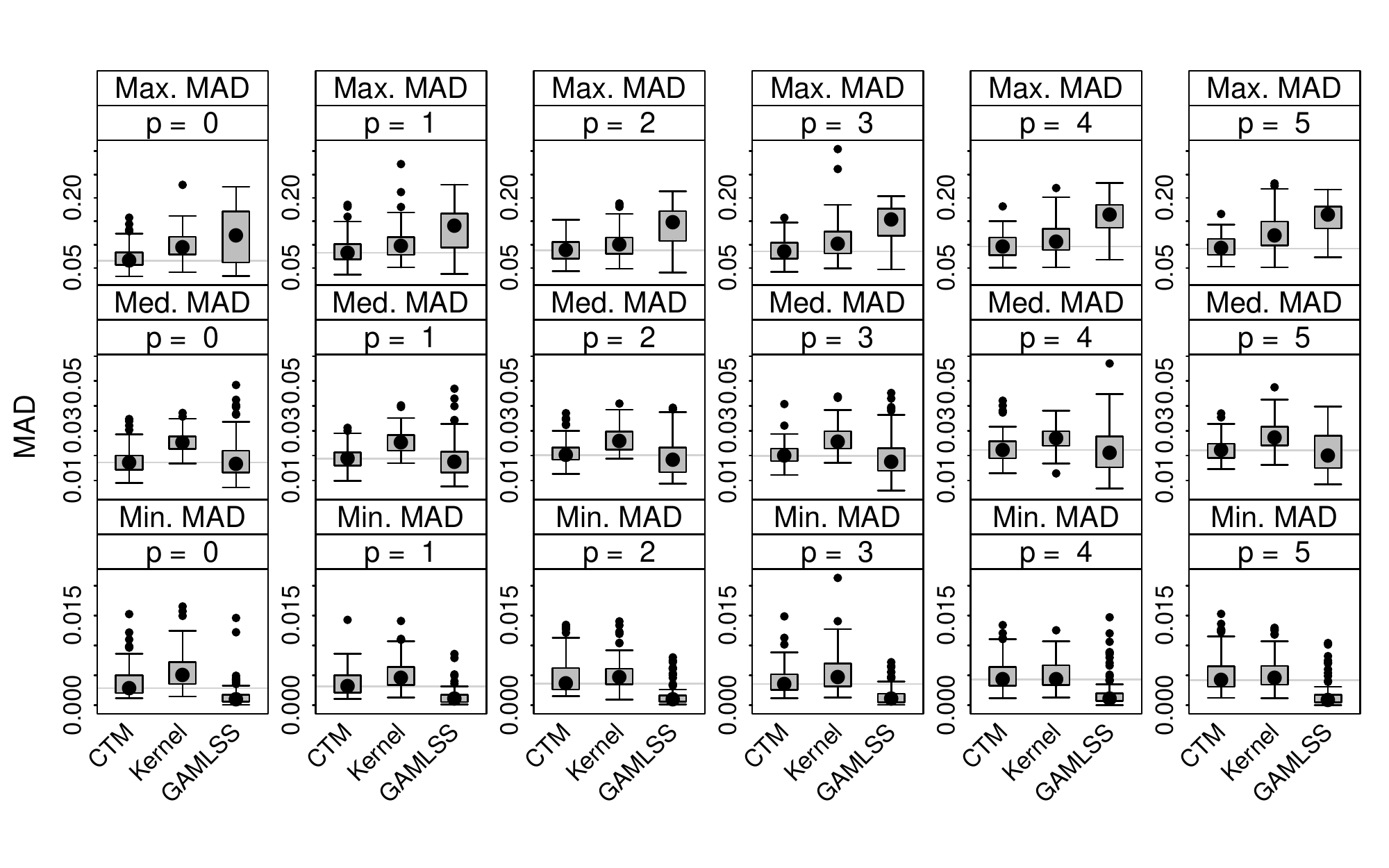}
\end{center}
\caption{Empirical Evaluation. Minimum, median, and maximum of the mean absolute deviation (MAD)
         between true and estimated probabilities for conditional transformation
         models (CTM), non-parametric kernel distribution function estimation (Kernel),
         and generalised additive models for location, scale, and shape (GAMLSS) for
         $100$ random samples.
         Values on the ordinate can be interpreted as absolute differences of probabilities.
         The grey horizontal lines correspond to the median of CTM.
         \label{sim-plot}}
\end{sidewaysfigure}

Figure~\ref{sim-plot} shows the empirical distributions of the minimum,
median and maximum MAD for the three
competitors.  For $p = 0$, GAMLSS and conditional transformation models
perform on par with respect to the median MAD, although GAMLSS shows a
somewhat larger variability.  The median MAD is slightly smaller than $0.02$
for both procedures, which indicates that the true conditional distribution
function can be fitted precisely.  The maximal MAD is smallest for
conditional transformation models and can be quite large for GAMLSS.  In
contrast, for some configurations of the explanatory variables, GAMLSS seems
to offer better estimates with respect to the minimal MAD.  The kernel
estimator leads to the largest median MAD values but seems more robust than
GAMLSS with respect to the maximal MAD.  These results are remarkably robust
in the presence of up to five non-informative explanatory variables,
although of course the MAD increases with $p$.

The general theme that GAMLSS on average performs as well as conditional
transformation models in the special case of model~(\ref{simmod}) but is
associated with a larger variability might be explained by the
independent estimation of the functions for the expectation and variance, \ie
GAMLSS does not ``know'' that the varying coefficient term and the variance
term are actually the same.  The inferior performance of the kernel
estimator might be explained by the technical difficulties associated with
bandwidth choice.  The tuning parameters for the two boosting approaches are
easier to choose.  Our general impression is that the kernel-estimated
conditional distribution functions are more erratic than the smooth
functions obtained with boosting for conditional transformation models
(analysis of simulation data not shown here).  
%%As an illustration of this phenomenon, we fitted both models for the Italian
%%gross domestic product data analysed by \cite{Hayfield_Racine_2008} and
%%depict corresponding quantiles in Figure~\ref{Italy-plot}.  The number of
%%boosting iterations for the conditional transformation model was determined
%%by the bootstrap and the bandwidths for the kernel estimator by
%%cross-validation.  The quantile functions obtained from the conditional
%%transformation model are smoother compared to the functions obtained from
%%the kernel distribution estimator.

%%\input{ex_italy}

\begin{figure}[t]
\begin{center}
\includegraphics{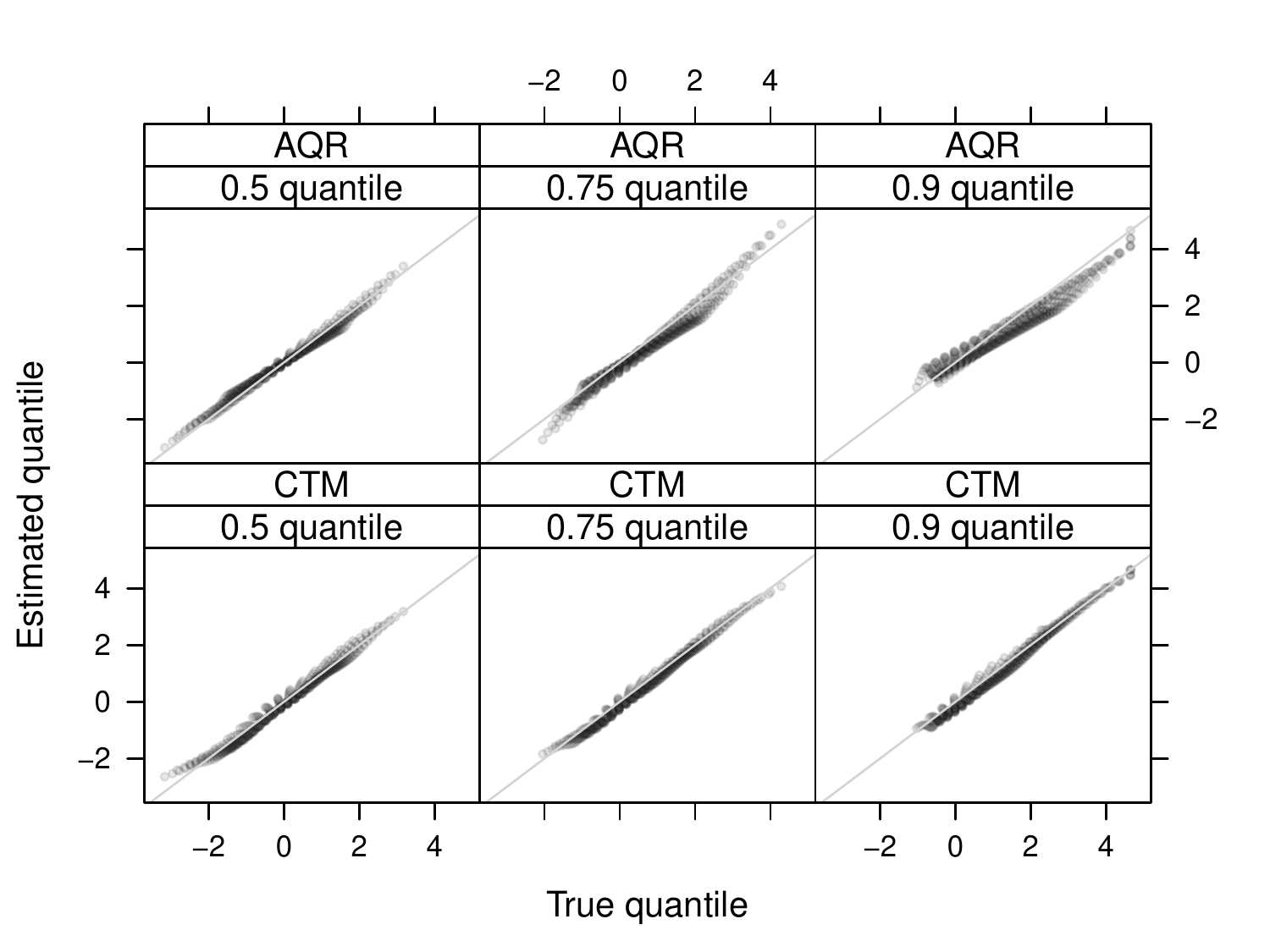}
\caption{Comparison of conditional transformation models and additive quantile regression.
         Scatterplot of true versus estimated quantiles obtained from one conditional transformation
         model (CTM) and from three additive quantile regression (AQR) models fitted to 
         $200$ observations drawn from the heteroscedastic varying coefficient 
         model (\ref{simmod}). \label{qsim-plot}}
\end{center}
\end{figure}

Since conditional transformation models are also an alternative to quantile
regression models, it would be interesting to compare both approaches.  At
this point, it is important to recall that the two models assume additivity
of the effects of $X_1$ and $X_2$, however, on different scales as explained
in Section~\ref{sec:ctm}.  Consequently, the heteroscedastic varying
coefficient model (\ref{simmod}) cannot be fitted in a straightforward way
using standard linear or additive quantile regression.  Though, the
estimation problem can be slightly reformulated by describing the
$\tau$-quantile of $\Yx$ as the sum of a varying coefficient term
$r_1(x_1)x_2$ and a smooth function $r_2(x_1)$.  This model, implemented
using the boosting approach to additive quantile regression with varying
coefficients introduced by \cite{Fenske_Kneib_Hothorn_2011}, allows the
estimation of conditional $\tau$-quantiles.  We fitted three such quantile regression 
models (for $\tau = 0.5, 0.75, 0.9$) to a
sample of size $N = 200$ from model (\ref{simmod}) and determined the
optimal number of boosting iterations by the out-of-bootstrap empirical risk
of the check function. To give an impression, we
compare these estimated $\tau$-quantiles with the corresponding conditional
quantiles obtained by inverting the estimated conditional distribution
function from a conditional transformation model.  Figure~\ref{qsim-plot}
displays scatterplots of the true conditional quantiles over a grid of $x_1$
and $x_2$ values with the corresponding estimated quantiles derived from one
conditional transformation model and the three additive quantile regression
models for $\tau = 0.5, 0.75, 0.9$, the latter models including the
varying coefficient term.  It seems that, in this example, both approaches
recover the true quantiles equally good.

\section{Discussion} \label{sec:dis}

In his book \textit{Quantile Regression}, \cite{Koenker_2005} puts transformation
models in the ``twilight zone of quantile regression'' and suggests that
estimating conditional distribution functions by means of transformation
models might be an alternative to the direct estimation of conditional
quantile functions.  We undertook the ``worthwhile exercise''
\citep[][Section 8.1.1.]{Koenker_2005} and developed a semiparametric
framework for the estimation of conditional distribution functions by
conditional transformation models that allows higher moments of the
conditional distribution to depend on the explanatory variables.

Because the empirical risk function (\ref{mod:risk}) is equivalent to
well-established risk functions for binary data, there are many potentially
interesting algorithms that can be used to fit conditional transformation
models, although dependent observations have to be dealt with.  We chose a
component-wise boosting approach mainly because of its divide-and-conquer
strategy, which allows a very efficient fitting of base-learners that depend
on the response and on one or more explanatory variables at the same time via
linear array models.  In addition, the algorithm is general, and
different base-learners appropriate for the model at hand can be
easily implemented as illustrated in Section~\ref{sec:app}.  
Also very
attractive are the model choice properties, as for example illustrated by
means of the Boston Housing data in Section~\ref{sec:app}. The
theoretical and empirical properties investigated in Sections~\ref{sec:con} 
and \ref{sec:eval} indicated that the method 
is applicable to more high-dimensional situations as well. While boosting
became popular owing to its success in fitting simple models under challenging
circumstances -- especially linear or additive models for 
high-dimensional explanatory variables -- the attractiveness of this class of
algorithms for fitting challenging models in simple circumstances has been
only rarely recognised.  Exceptions are \cite{Ridgeway_2002} and
\cite{Sexton_Laake_2012}, who study boosting algorithms for fitting density
functions.  \cite{Lu_Li_2008,Schmid_Hothorn_2008} and
\cite{Schmid_Hothorn_Maloney_2011} proposed boosting algorithms for
transformation models that treat the transformation function $h_Y$ as
a nuisance parameter.  In the same model framework, \cite{Tutz_Groll_2012}
propose a likelihood-boosting approach for fitting cumulative and sequential
models for ordinal responses.  

Boosting algorithms for estimating conditional
quantiles by minimising the check function have been introduced by
\cite{Kriegler_Berk_2010,Fenske_Kneib_Hothorn_2011}, and \cite{Zheng_2012}. 
The computation of prediction intervals based on pairs of such models 
is rather straightforward \citep{Mayr_Hothorn_Fenske_2012}. Our approach to
the estimation of the conditional distribution function has the advantage
that one model fits the whole distribution, which can then be used to derive
arbitrary functionals from. The quantile score representation of
the continuous ranked probability score \citep[see][]{Gneiting_Ranjan_2011}
might be a basis to develop a boosting technique similar to the one described
in this paper for the estimation of full conditional quantile functions. The
main difference between transformation and quantile regression models that
we have to keep in mind is that additivity is assumed on two different
scales. From a practical point of view,
diagnostic tools to assess on which of these scales it is more appropriate 
to assume an additive model would be very important.

The applications presented in Section~\ref{sec:app} showed that conditional
transformation models are generic, and one can, by choosing
appropriate base-learners, fit models that are specific to the problem at
hand. An empirical evaluation showed that the estimated conditional
distribution functions are on average as good as the estimates obtained from
a parametric approach (GAMLSS) that relies on more assumptions. In
comparison to non-parametric kernel distribution estimators, conditional
transformation models are more adaptable, for example to spatial or
temporal data. The performance of the semiparametric models compared to that of the
non-parametric competitor was considerably better at the small price of
the assumption of additivity of the transformation function.

It will be interesting to further study conditional transformation models 
with respect to the following extensions. Instead of making assumptions
about the quantile function $Q$ representing the error distribution, 
it would be possible to fit the corresponding distribution function by
techniques introduced for single index models \citep{Tutz_Petry_2012}. 
To reduce the complexity, one
could restrict the partial transformation functions $h_j$ to linear
functions, \ie only the first two moments are allowed to be influenced by
the explanatory variables. Furthermore, when the response $Y = (Y_1, Y_2)$ is bivariate,
the bivariate conditional distribution function $\Prob(Y_1 \le \upsilon_1,
Y_2 \le \upsilon_2 | \mX = \xvec)$ can be estimated by minimising the risk
function
\begin{eqnarray*}
  \int \int \rho\{((y_1 \le \upsilon_1 \wedge y_2 \le \upsilon_2), \xvec), h(\upsilon_1, \upsilon_2 |
\xvec)\}
  \,d\mu(\upsilon_1) \,d\mu(\upsilon_2) \,d \hatPYX(y_1, y_2, \xvec)
\end{eqnarray*}
in $h$. It might also be interesting to allow only certain moments to depend
on certain explanatory variables. For example, a partially conditional
transformation model of the form
\begin{eqnarray*}
h(\upsilon | \xvec) = \sum_{j = 1}^{J_1} h_j(\upsilon | \xvec) + 
\sum_{j = J_1 + 1}^{J} h_j(1 | \xvec)
\end{eqnarray*}
describes the conditional expectation $\sum_{j = J_1 + 1}^{J} h_j(1 | \xvec)$
of a transformation $\sum_{j = 1}^{J_1} h_j(\Yx | \xvec)$.

Researchers pay most attention to the conditional distribution function 
in applications where the response is a survival time or has been
censored owing to other reasons.  In survival analysis, it is common to deal
with the conditional survivor function $S(\Yx | \xvec) = 1 - F(\Yx | \xvec)$ for
survival time $Y$.  In the seemingly simpler situation of an uncensored
continuous response, many data analysts focus on the conditional mean and
do not bother with the conditional distribution function at all. Most of the
literature on transformation models therefore deals with the censored case. 
The easiest way to deal with right-censoring is to minimise the risk
function weighted by inverse probability of censoring weights, as in the
generalised estimating equations approach by \cite{Cheng_Wei_Ying_1995},
where parameters of a linear transformation model are estimated by the root of 
a V-statistic defined by all binary indicators $I(Y_i \ge Y_\imath),
i,\imath = 1, \dots, N$. Accelerated failure time models fitted by boosting
of an inverse probability of censoring weighted risk function have been
described by \cite{Hothorn_Buehlmann_Dudoit_2006}, and future research
awaits the investigation of the performance of conditional transformation models
under censoring.

\section*{Computational Details}

Conditional transformation models were fitted using an implementation
of component-wise boosting in package \pkg{mboost} \citep[version~2.1-2,][]{pkg:mboost}. Package
\pkg{gamboostLSS} \citep[version~1.0-3,][]{pkg:gamboostLSS} was used to fit GAMLSS models and kernel
distribution estimation was performed using package \pkg{np} \citep[version~0.40-13,][]{pkg:np}.
Linear quantile regression was computed using package \pkg{quantreg} \citep[version~4.79,][]{pkg:quantreg}.
All computations were performed using \textsf{R} version 2.13.2 \citep{R2.13.2}.

Throughout Section~\ref{sec:app}, we used the loss function defined by
$\rho_\text{bin}$ and modelled non-linear functions by cubic $B$-spline
bases with $20$ equi-distant knots.  For further computational details we
refer the reader to the \textsf{R} code that implements the analyses
presented in Sections~\ref{sec:app} and \ref{sec:eval}, which is available
in an experimental \textsf{R} package \pkg{ctm} at
\url{http://R-forge.R-project.org/projects/ctm}.  The results presented in
this paper can be reproduced using this package, except for the birth weight
data, which are not publically available.

\section*{Acknowledgements}

We would like to thank Gareth O. Roberts, an anonymous associate editor and
referee for their careful review of an initial version of this paper.
Andreas Mayr implemented the \texttt{Normal()} family for
\textbf{gamboostLSS} on our request, Achim Zeileis helped finding the right
colors for Figure~\ref{india_qplot} and Ronald Schild provided us with the birth weight data.
We are indepted to Paul Eilers, Tilman Gneiting and Roger Koenker for their comments on 
a draft version and thank Karen A.~Brune for improving the language.
Financial support by Deutsche Forschungsgemeinschaft (grant HO 3242/4-1) is
gratefully acknowledged.

%% \clearpage

\bibliography{refs} %%%,packages}

\newpage

\begin{appendix}

\section*{Appendix}

\subsection*{Proofs}

\begin{proof}{Proof of Lemma 1}

Convexity: If the loss function $\rho$ is convex in                         
its second argument, so is the loss function $\ell$
\begin{eqnarray*}
\ell((Y, \mX), \alpha h + (1 - \alpha) g) \le \alpha \ell((Y, \mX), h) + 
  (1 - \alpha) \ell((Y, \mX), g),
\end{eqnarray*}
with $g(\cdot | \xvec): \RR \rightarrow \RR$ being a
monotone increasing transformation function and $\alpha \in [0, 1]$,
because of the convexity of $\rho$ and the monotonicity and linearity
of the Lebesgue integral.

Population Minimisers: Let $f$ denote the density of $F$. 
With iterated expectation we have
\begin{eqnarray*}
\ExYX \ell((Y, \mX), h) 
& = & \int \int \int \rho((y \le \upsilon, \xvec), h(\upsilon | \xvec)) 
        \,d\mu(\upsilon) \,d \PYXx(y) \,d\PX(\xvec) \\
& = & \int \int 
  \underbrace{\int \rho((y \le \upsilon, \xvec), h(\upsilon | \xvec)) 
                \,d \PYXx(y)}_{=:A_{\upsilon, \xvec}(h(\upsilon | \xvec))} \,d\mu(\upsilon) \,d\PX(\xvec)
\end{eqnarray*}
and the risk is minimal when $A_{\upsilon, \xvec}(h(\upsilon | \xvec))$ is 
minimal for the scalar $h(\upsilon | \xvec)$ for all 
$\upsilon$ and $\xvec$, \ie when
\begin{eqnarray*}
0 
& \stackrel{!}{=} & \frac{\partial A_{\upsilon, \xvec}(h(\upsilon | \xvec))}
                         {\partial h(\upsilon | \xvec)} \\
& = & \int \frac{\partial}{\partial h(\upsilon | \xvec)} 
  \rho((y \le \upsilon, \xvec), h(\upsilon | \xvec)) \,d \PYXx(y) \\
& \stackrel{\rho = \rho_\text{sqe}}{=} & \int (I(y \le \upsilon) - 
  F(h(\upsilon | \xvec)) f(h(\upsilon | \xvec)) \,d \PYXx(y) \\
& = & f(h(\upsilon | \xvec)) \left\{ \int I(y \le \upsilon) \,d \PYXx(y) -
  F(h(\upsilon | \xvec)) \right\} \\
& = & f(h(\upsilon | \xvec)) \left\{ \Prob(Y \le \upsilon | \mX = \xvec) -
  F(h(\upsilon | \xvec)) \right\}
\end{eqnarray*}
which for $f(h(\upsilon | \xvec)) > 0$ is zero for
$h(\upsilon | \xvec) = F^{-1}(\Prob(Y \le \upsilon | \mX = \xvec))$.
Similar, for $\rho = \rho_\text{bin}$ the term
\begin{eqnarray*}
0
& \stackrel{!}{=} & \frac{\partial A_{\upsilon, \xvec}(h(\upsilon | \xvec))}
                         {\partial h(\upsilon | \xvec)} \\ 
& = & \int -\left\{\frac{I(y \le \upsilon)}{F(h(\upsilon | \xvec))} 
  f(h(\upsilon | \xvec)) - 
  \frac{1 - I(y \le \upsilon)}{1 - F(h(\upsilon | \xvec))}
  f(h(\upsilon | \xvec))\right\} \,d \PYXx(y) \\
& = & f(h(\upsilon | \xvec)) \left\{ 
  \frac{\int 1 - I(y \le \upsilon) \,d \PYXx(y) }{1 - F(h(\upsilon | \xvec))} - 
  \frac{\int I(y \le \upsilon) \,d \PYXx(y)}{F(h(\upsilon | \xvec))} \right\} \\
& = & f(h(\upsilon | \xvec)) \left\{ 
  \frac{1 - \Prob(Y \le \upsilon | \mX = \xvec)}{1 - F(h(\upsilon | \xvec))}  - 
  \frac{\Prob(Y \le \upsilon | \mX = \xvec)}{F(h(\upsilon | \xvec))} \right\}
\end{eqnarray*}
is zero for $h(\upsilon | \xvec) = F^{-1}(\Prob(Y \le \upsilon | \mX =
\xvec))$ when $f(h(\upsilon | \xvec)) > 0$.

For the absolute error, note that
\begin{eqnarray*}
\rho_\text{abe}((Y \le \upsilon, \mX), h(\upsilon | \mX)) =  
  I(Y \le \upsilon)\{1 - F(h(\upsilon | \mX))\} + 
  \{1 - I(Y \le \upsilon)\} F(h(\upsilon | \mX))
\end{eqnarray*}
and thus
\begin{eqnarray*}
A_{\upsilon, \xvec}(h(\upsilon | \xvec)) 
& = & \int \rho_\text{abe}((y \le \upsilon, \xvec), h(\upsilon | \xvec)) 
  \,d \PYXx(y) \\
& = & \int   I(Y \le \upsilon)\{1 - F(h(\upsilon | \mX))\} + 
  \{1 - I(Y \le \upsilon)\} F(h(\upsilon | \mX)) \,d \PYXx(y) \\
& = & \left\{1 - F(h(\upsilon | \mX))\right\} \Prob(Y \le \upsilon | \mX = \xvec) + 
  F(h(\upsilon | \mX))\{1 - \Prob(Y \le \upsilon | \mX = \xvec)\}
\end{eqnarray*}
This expression attains its minimal value of 
$\Prob(Y \le \upsilon | \mX = \xvec)$ for 
$\Prob(Y \le \upsilon | \mX = \xvec) \le 0.5$ 
when $F(h(\upsilon | \mX)) = 0$. For 
$\Prob(Y \le \upsilon | \mX = \xvec) > 0.5$, the minimum 
$1 - \Prob(Y \le \upsilon | \mX = \xvec)$
is attained when $F(h(\upsilon | \mX)) = 1$. 
Thus, absolute error will lead to too extreme
estimated values of $h$ and corresponding conditional distribution functions.
\end{proof}

\begin{proof}{Proof of Theorem \ref{th1}}

We use a modified argument of a proof presented in Section 12.8.2.~in
\cite{Buehlmann_Geer_2011}. Formally, we can write  
\begin{eqnarray*}
& &I(Y_i \le \upsilon_\imath) = h_{\gammavec_{0,N}}(\upsilon_\imath|\mX_i) 
+ \eps_{i\imath},\\
& &\eps_{i\imath} = I(Y_i \le
  \upsilon_\imath) - h_{\gammavec_{0,N}}(\upsilon_\imath|\mX_i)\ (i=1,\ldots ,N,\
\imath=1,\ldots ,n). 
\end{eqnarray*}
The errors $\eps_{i\imath}$ have reasonable properties, as discussed in
(\ref{err}) below.  

There are two issues that need to be addressed. 
First, we define the inner products of functions $h$ and $g$
$$(h,g)_{n,N,\Ex} = n^{-1} \sumimath \Ex[h(\upsilon_\imath | \mX) 
                                         g(\upsilon_\imath | \mX)]$$ 
and 
$$(h,g)_{n,N} = n^{-1} N^{-1} \sumi \sumimath h(\upsilon_\imath | \mX_i) 
                                              g(\upsilon_\imath | \mX_i).$$ 
The proof in \cite{Buehlmann_Geer_2011} can then
be used with the scalar product $(h,g)_{n,N}$.   

Secondly, for controlling the probabilistic part of the proof, we need to
show that the analogue of formula (12.26) in \cite{Buehlmann_Geer_2011} holds. 
This translates to deriving a bound for
\begin{eqnarray*}
\max_{j,k_0,k_1} (b_{0,k_0} b_{j,k_1},\eps)_{n,N} =
\max_{j,k_0,k_1} (nN)^{-1} \sumi \sumimath 
b_{0,k_0}(\upsilon_\imath) b_{j,k_1}(X_j) \eps_{i\imath}.
\end{eqnarray*}
Because $h_{\gammavec_{0,N}}$ is the projection of $I(Y \le \upsilon_\imath)\
(\imath=1,\ldots,n)$ onto the basis functions \\
$b_{0,k_0}(\upsilon_\imath) b_{j,k_1}(X_j) \imath=1,\ldots ,n)$
with respect to the $\|\cdot\|_{n,N,\Ex}$-norm (see (\ref{proj})), and owing
to the definition of $\eps_{i\imath}$, we have 
\begin{eqnarray}\label{err}
n^{-1} \sumimath \Ex[\eps_{i\imath} b_{0,k_0}(\upsilon_\imath)
b_{j,k_1}(X_j)]
= 0\ \forall j,k_0,k_1. 
\end{eqnarray}
Therefore,
\begin{eqnarray*}
& &(nN)^{-1} \sumi \sumimath b_{0,k_0}(\upsilon_\imath)
b_{j,k_1}(X_j) \eps_{i\imath} = N^{-1} \sumi Z_i(j,k_0,k_1),\\
& &\Ex[Z_i(j,k_0,k_1)] = 0.
\end{eqnarray*}
Furthermore, owing to the boundedness assumption in (A1),
$\|Z_i(j,k_0,k_1)\|_{\infty} \le C_1$ for some constant $C_1 <
\infty$ $\forall i,j,k_0,k_1$. Applying Hoeffding's inequality, for
independent (but not necessarily identically distributed) random variables
\citep[Lem. 3.5]{Geer_2000} and using the union bound, we obtain 
\begin{eqnarray*}
\max_{j,k_0,k_1} (b_{0,k_0} b_{j,k_1},\eps)_{n,N} =
O_P\left(\sqrt{\log(J_N K_{0,N} K_{1,N})/N}\right).
\end{eqnarray*}
This, together with the proof from Section 12.8.2 in \cite{Buehlmann_Geer_2011}
completes the proof of Theorem \ref{th1}.
\end{proof}

\subsection*{Gradients}

Gradients for different loss functions $\rho$ and arbitrary absolute
continuous distribution functions $F$ with density function $f$:
\begin{eqnarray*}
U_{i\imath} & \stackrel{\rho = \rho_\text{bin}}{=} & \left\{
  \frac{I(Y_i \le \upsilon_\imath)}{F\left(\hat{h}^{[m]}_{i\imath}\right)}
  - \frac{1 - I(Y_i \le \upsilon_\imath)}{1 + F\left(\hat{h}^{[m]}_{i\imath}\right)}
  \right\} f\left(\hat{h}^{[m]}_{i\imath}\right) \\
U_{i\imath} & \stackrel{\rho = \rho_\text{sqe}}{=} &
  \left\{I(Y_i \le \upsilon_\imath) -
F\left(\hat{h}^{[m]}_{i\imath}\right)\right\}
  f\left(\hat{h}^{[m]}_{i\imath}\right) \\
U_{i\imath} & \stackrel{\rho = \rho_\text{abe}}{=} &
  \left\{I(Y_i \le \upsilon_\imath) - (1 - I(Y_i \le
\upsilon_\imath))\right\}
  f\left(\hat{h}^{[m]}_{i\imath}\right)
\end{eqnarray*}

\end{appendix}

\end{document}